\documentclass[preprint,floatfix
,aps,prd,superscriptaddress]{revtex4}

\usepackage[utf8]{inputenc}
\usepackage{graphicx}
\usepackage{tikz}
\usepackage{amssymb,amsmath,mathtools,comment,hyperref}
\usepackage{appendix}

\usepackage{extarrows}
\usepackage{relsize}

\usepackage{bm}
\usepackage{mathrsfs}

\newcommand{\ket}[1]{\left|#1\right\rangle}
\newcommand{\bra}[1]{\left\langle #1 \right|}

\newcommand{\mo}{\mathcal{O}}
\newcommand{\T}{\theta}
\newcommand{\D}{\Delta}
\newcommand{\s}{\sin}
\newcommand{\C}{\cos}

\newcommand{\rs}[1]{\rho^{\rm #1}}

\begin{document}

\title{Quantum optics meets black hole thermodynamics
 \\
via conformal quantum mechanics:
\\
II. Thermodynamics of acceleration radiation}

\author{A. Azizi}
\affiliation{Institute for Quantum Science and Engineering, Texas A\&M University, College Station, Texas, 77843, USA}
\author{H. E. Camblong}
\affiliation{Department of Physics and Astronomy, University of San Francisco, San Francisco, California 94117-1080, USA}
\author{A. Chakraborty}
\affiliation{Department of Physics, University of Houston, Houston, Texas 77024-5005, USA}
\author{C. R. Ord\'{o}\~{n}ez}
\affiliation{Department of Physics, University of Houston, Houston, Texas 77024-5005, USA}
\affiliation{Department of Physics and Astronomy, Rice University, MS 61, 6100 Main Street, Houston, Texas 77005, USA.}
\author{M. O. Scully}
\affiliation{Institute for Quantum Science and Engineering, Texas A\&M University, College Station, Texas, 77843, USA}
\affiliation{Baylor University, Waco, TX 76706, USA}
\affiliation{Princeton University, Princeton, New Jersey 08544, USA}

\date{\today}
\begin{abstract}
The thermodynamics of ``horizon brightened acceleration radiation'' (HBAR),
due to a random atomic cloud freely falling into a black hole in a Boulware-like vacuum, is shown to 
mimic the thermodynamics of the black hole itself. 
The thermodynamic framework is developed in its most general form via a quantum-optics master equation, including rotating (Kerr) black holes and for any set of initial conditions of the atomic cloud. 
The HBAR field exhibits thermal behavior at the Hawking temperature and an area-entropy-flux relation that resembles the Bekenstein-Hawking entropy. In addition, this general approach reveals:
(i) the existence of an {\it HBAR-black-hole thermodynamic correspondence\/} that explains the HBAR area-entropy-flux relation; 
(ii) the origin of the field entropy from the near-horizon behavior, via conformal quantum mechanics (CQM).

\end{abstract}
\maketitle

\section{Introduction}
\label{sec:introduction}
Two of the central pillars of black hole thermodynamics~\cite{BH-thermo_reviews_1,BH-thermo_reviews_2}
are 
the Bekenstein-Hawking entropy $S_{\rm BH}$~\cite{bekenstein1972, bekenstein1973, bekenstein1974}
and the Hawking radiation effect~ \cite{hawking74,hawking75}, along with the Hawking temperature $T_{H}$.
These results
appear to be universal properties of any theory combining quantum physics with gravitation.
While their true origin is still elusive after almost five decades, 
progress has been made with string theory~\cite{BH_string} and loop quantum gravity~\cite{BH_loop}.
Moreover, all approaches suggest that the origin of the thermodynamics is related to the event horizon~\cite{BH-thermo_reviews_2}, possibly in the form of a conformal field theory~\cite{strominger:98,carlip:near_horizon,solodukhin:99,sen01}.
Furthermore, due to the equivalence principle, related results have been identified in accelerated systems in the form of the Fulling-Davies-Unruh effect and the associated Unruh temperature~\cite{unruh76,fulling76,davies77}.
Additional insights into these profound concepts are of great interest; thus, in this paper, we probe deeper into some nontrivial connections between the thermodynamics of black holes and acceleration radiation. 

We use the foundational results of the first article in this series~\cite{HBAR_part-I}
to fully develop the thermodynamics of ``horizon brightened acceleration radiation'' (HBAR) generated by a 
an atomic cloud in free fall into a black hole in a Boulware-like vacuum, with random injection times.
These results generalize to rotating black holes the quantum optics approach of Ref.~\cite{scully2018}.
 Most importantly, as in the preceding article~\cite{HBAR_part-I}, conformal quantum mechanics (CQM)
 is shown to fully drive the thermodynamic behavior~\cite{camblong2005,nhcamblong-sc,camblong2013,camblong2020,azizi2021}.
 Such a symmetry-based approach is appealing as it supports the notion that conformal invariance may play a crucial role in a deeper understanding of black hole thermodynamics~\cite{strominger:98,carlip:near_horizon,solodukhin:99,sen01}, and ultimately, in a theory of quantum gravity. 
In essence, the ensuing thermodynamic framework relies on the primary thermal properties that consist of the Hawking temperature and thermality via a detailed-balance Boltzmann factor---these are common to both HBAR and black hole thermodynamics. 
With these tools, this paper establishes the existence of formally identical thermodynamic functional relationships, which we describe as the HBAR-black-hole thermodynamic correspondence and include the HBAR area-entropy-flux relation. In particular, the HBAR entropy flux
is proportional to the rate of change in horizon surface area due to the photon emission, with the critical  proportionality factor that is exactly $1/4$.

\begin{figure}[t]
    \centering
    \includegraphics[width=1.0\linewidth]{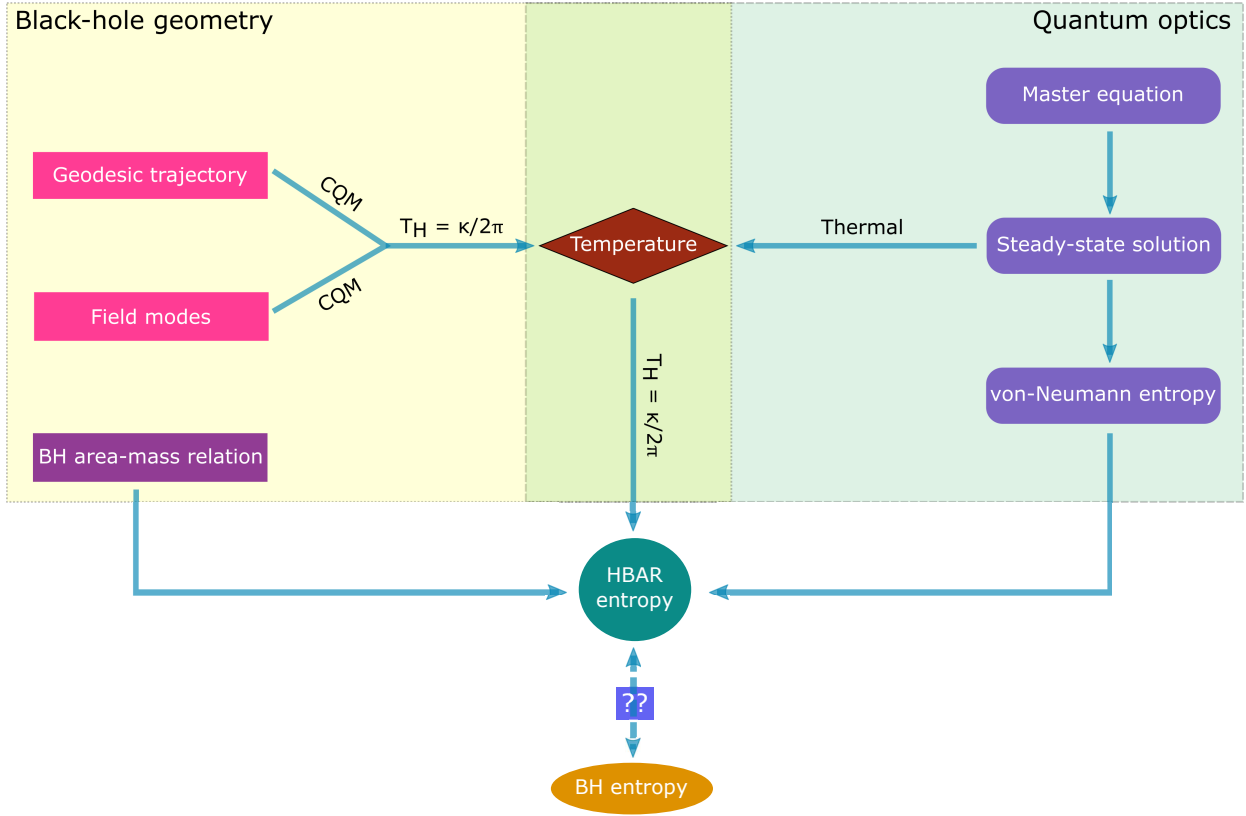}
    \caption{Logical flow of basic concepts in this article. On the right hand side we have the elements of quantum optics that leads to a thermal steady state. The black-hole (BH) geometry on the left side assigns the temperature of the thermal steady state to be Hawking temperature via near-horizon CQM. This assignment of temperature connects the BH aspects of the problem with the quantum optics methodology. What remains to be explored is exactly how this relates to the BH entropy on a deeper level.}    \label{fig:flow-chart_logic}
\end{figure}

The organization of this article is as follows.
It involves two intertwining
tracks, properly addressed in the different sections: 
the physics of the background gravitational field and the statistical quantum-optics approach leading to the thermodynamics through the master equation for the reduced field density matrix.
The logical progression of concepts and their interrelationship are outlined in Fig.~\ref{fig:flow-chart_logic}, with the details on the master equation briefly summarized
 from the extensive treatment of Ref.~\cite{HBAR_part-I}.
In Section~\ref{sec:setup}, the basic concepts for both tracks are introduced: 
the geometry, the interactions, and a review of the master equation from Ref.~\cite{HBAR_part-I}.
In Sec.~\ref{sec:nh-CQM_field-geodesics}, we show that the physics of the scalar field in the gravitational background near the event horizon is governed by CQM; both the near-horizon field equations and geodesics are derived.
  In Sec.~\ref{sec:conformal_steady_state}, we examine the implications of the governing physics of CQM,
   both in terms of the gravitational background and the master equation; in particular, this includes
   a derivation of the Planck form of the atom's probability of emission of photons, the existence of detailed balance Boltzmann factor associated with the Hawking temperature, and 
  a characterization of the thermal nature of the field state via the master equation.
 These buildup of concepts culminates in Sec.~\ref{sec:HBAR-thermo}, with a thorough analysis of the HBAR thermodynamics leading to the HBAR-black-hole thermodynamic correspondence,
 and including the HBAR entropy flux formula of Ref.~\cite{scully2018}, with a general proof of its conformal nature. 
Concluding remarks are given in Sec.~\ref{sec:conclusion}, and followed by the appendices, which include 
a summary of the Kerr geodesics (\ref{app:Kerr-geodesics})
and the technicalities of Kerr-geometry modes and vacuum states (\ref{app:vacuum_modes}).

\section{Basic concepts: Kerr geometry, atom-field-gravity interactions, and field master equation}
\label{sec:setup}
\subsection{Kerr geometry}
The spacetime geometry provides the gravitational
background where the atom-field interactions take place. In this paper, we will focus on the geometry due to non-extremal Kerr black holes, as these are of current interest and broad generality. 
as representatives of the rotating class of black holes in 4D.
Specifically, the Kerr metric describes the spacetime geometry that is the exact vacuum solution of the Einstein
 general relativistic field equations in 4D in the presence of a black hole of mass $M$ and angular momentum $J$. 
In Boyer-Lindquist coordinates $(t,r,\theta,\phi)$, the metric admits several equivalent expressions; 
in its most basic form, in geometrized units $c= G= 1$, it is given by
\begin{equation}
ds^2 =
-\frac{(\Delta - a^2 \s^2 \T)}{\rho^2} dt^2
- \frac{4 M r}{\rho^2} a \s^2 \T dt d \phi
+\frac{\rho^2}{\Delta}dr^2 +\rho^2d\T^2
+ \frac{\Sigma^2}{\rho^2} \s^2 \T d \phi^2 
\, 
\label{eq:Kerr1}
\end{equation}
where 
the Kerr parameter
$a=J/M$, is the
angular momentum per unit mass,
and the symbols $\D$, $\rho$, and $\Sigma$ are defined as
\begin{equation}
\begin{aligned}
\D&= r^2 -2Mr + a^2
\;  \;  , \; \; \; \; \; 
\rho^2 = r^2 + a^2 \C^2\T
\; , \\
\Sigma^2 &= \left( r^2 +a^2 \right) \rho^2 + 2 M r a^2 \, \s^2 \T = (r^2 +a^2)^2 - \D a^2 \s^2 \T
\; .
\end{aligned}
\end{equation}
The analysis of this paper is similarly valid for the Kerr-Newman geometry, which has the same structural form 
form with an additional black hole electric charge $Q$, which shows up as a modification
$\D = r^2 -2Mr + a^2 +Q^2$, leading to a replacement $a^2 \rightarrow a^2 + Q^2$ in
Eq.~(\ref{eq:r_pm}) and ensuing equations.

An alternative form of the Kerr metric,
\begin{equation}
ds^2  = -\frac{\D \rho^2}{\Sigma^2} dt^2 
 + \frac{\rho^2}{\Delta}dr^2 + \rho^2d\T^2    
+ \frac{\Sigma^2}{\rho^2} \s^2 \T \left( d \phi - \varpi dt \right)^2
\; ,
\label{eq:Kerr3} 
\end{equation}
can be derived by absorbing the off-diagonal term $g_{t\phi}$ in a shift of the angular coordinate $\phi$, such that
 \begin{equation}
 \varpi = - \frac{ g_{t\phi} }{ g_{\phi \phi} }
 \, 
 \label{eq:Kerr-angular-velocity}
 \end{equation}
 is interpreted as a position-dependent angular velocity~\cite{Landau-Lifshitz_CTFs}. 
Equation~(\ref{eq:Kerr-angular-velocity}) is particularly useful to understand the physics 
as the event horizon is approached, as we will further analyze in Sec.~\ref{sec:nh-CQM_field-geodesics}.

The roots of the equation 
$g^{rr} = 0$, which amounts to 
$\D=0$, give the locations of the outer and inner event horizon,
\begin{equation}
r_\pm = M\pm (M^2-a^2)^{1/2}
\label{eq:r_pm}
\; .
\end{equation}
In addition, 
 the ergosphere is the region between the outer event horizon and 
 the outer static limit, which is defined by the largest root of $g_{tt}=0$.
The non-extremal geometry that we study in this paper corresponds to the physical condition $M>a$, 
which amounts to  $\D'_{+} \equiv \D'(r_{+}) = r_+-r_- \neq 0$, and where 
the prime stands for the radial derivative.

The symmetries of the metric~(\ref{eq:Kerr1}) lead to the Killing vectors
\begin{equation}
\boldsymbol\xi_{(t)} = \partial_t 
\; \; , \; \; \; \; \;
\boldsymbol\xi_{(\phi)} = \partial_\phi
\; 
\label{eq:Killing-vectors}
\end{equation}
associated with the stationary and axisymmetric nature of the metric
(independence with respect to $t$ and $\phi$ respectively). In addition, the particular combination 
$\boldsymbol\xi = \boldsymbol\xi_{(t)} + \Omega_{H} \boldsymbol\xi_{(\phi)}$ 
defines a Killing vector with respect to which the horizon  $\mathcal{H}$ is a null hypersurface;
see the interpretation in Sec.~\ref{sec:nh-CQM_field-geodesics}, in and around Eq.~(\ref{eq:Killing-vector_tilde-t}). 
Three most relevant geometrical and physical properties of the black hole
are the horizon area $A$, the surface gravity $\kappa$, and the angular velocity $\Omega_{H}$.
The horizon area is
\begin{equation}
A= \int \sqrt{
g_{\theta \theta} g_{\phi \phi} } \, d \theta d \phi
= 4 \pi (r_{+}^{2}+a^{2}) 
\; .
\label{eq:area_Kerr}
\end{equation}
The surface gravity, geometrically defined from the Killing vector $\boldsymbol\xi $
 by the horizon value of 
$\kappa = -(\nabla_{\mu} \boldsymbol\xi_{\nu})( \nabla^{\mu} \boldsymbol\xi^{\nu})/2$ 
at $r_{+}$, takes the form 
\begin{equation}
\kappa= \frac{ \D'_{+} }{ 2 (r_{+}^2 +a^2) }
\; .
\label{eq:surface-gravity_Kerr}
\end{equation}
The angular velocity of frame dragging defines the angular velocity of the black hole as the limit 
 \begin{equation}
\Omega_{H} = \lim_{r \rightarrow r_{+}}  \varpi =  \frac{a}{r_+^2+a^2} =
 \frac{a}{2M r_{+}}
 \, .
 \label{eq:BH-angular-velocity}
\end{equation}

\subsection{Atom-field-gravity interactions}
In addition to the gravitational background of the Kerr geometry, our model 
 involves the two interacting systems: the atomic cloud and the quantum field, as discussed in Refs.~\cite{scully2018,camblong2020,azizi2021,HBAR_part-I}. 
 The basic interaction leading to acceleration radiation is modeled 
by means of a dipole coupling of a scalar field $\Phi$ with a freely falling atom.
The atoms are randomly injected and are freely fall into the Kerr black hole through a Boulware-like vacuum. 
For the Kerr metric, the definition of a Boulware-like vacuum is technically challenging because of the superradiant modes~\cite{Unruh-Starobinsky-rad,frolov};
this is discussed briefly in Ref.~\cite{azizi2021} and in Appendix~\ref{app:vacuum_modes} in greater detail.
One simple approach to overcome this difficulty is the introduction of a boundary to exclude the regions of asymptotic infinity, as discussed in Sec.~\ref{sec:conformal_steady_state}; the alternatives including the asymptotic regions are considered in Appendix~\ref{app:vacuum_modes}. Any such states qualify as Boulware-like and permit the generation of the HBAR radiation considered in this paper.

The quantization of the scalar field is expressed as 
\begin{equation}
    \Phi(
   \mathbf{r}
    ,t) = \sum_{\boldsymbol{s}} \left[
    a_{\boldsymbol{s}} \phi_{\boldsymbol{s}} (\mathbf{r}, t) + \mathrm{h.c.} 
    \right]
    \; ,
        \label{eq:field_expansion}
        \end{equation}
where $\mathrm{h.c.}$ stands for the Hermitian conjugate;
and $\mathbf{r} = (r, \theta,\phi)$ denotes the spatial Boyer-Lindquist coordinates for the metric~(\ref{eq:Kerr1}).
The lowering operator
$a_{\boldsymbol{s}}$ annihilates the Boulware-type vacuum, with corresponding field modes $\phi_{\boldsymbol{s}} $. The labeling of the field modes 
with the symbol ${\boldsymbol{s}}$ refers to the complete set of ``quantum numbers'' (including the mode frequency $\omega$); for the Kerr geometry in 3 spatial dimensions, this is
${\boldsymbol{s}} =  \{\omega,l,m\}$, where 
$\{l,m\}$ are the spheroidal number and the ``magnetic'' quantum number associated with angular momentum. 

The interaction of the field $\Phi$ of Eq.~(\ref{eq:field_expansion})
with a given two-level atom can be modeled as a weak-dipole coupling 
of strength $g$,
\begin{equation}
V_I(\tau) = 
g \, \Phi (\mathbf{r} (\tau),t(\tau)) \, \sigma
 \; ,
\label{eq:QO_interaction_potential}
\end{equation}
in which $\sigma = \sigma_- e^{-i\nu \tau} + \sigma_+ e^{i\nu \tau}$ is the operator
that causes atomic state transitions, with 
$\sigma_-$ being the corresponding atomic lowering operator.
Given at atom in its
ground state $\ket{b}$,
the coupling~(\ref{eq:QO_interaction_potential}) allows for the emission of a scalar photon with the simultaneous transition of the atom to its excited state
 $\ket{a}$; this process has a first-order perturbation probability amplitude
$-i g I_{{\mathrm e}, {\boldsymbol{s}} }$, where
$I_{{\mathrm e}, {\boldsymbol{s}} } =
\int d\tau \;\bra{1_{\boldsymbol{s}},a}V_I(\tau)\ket{0,b}$, and the field ground state $\ket{0}$
 and the state $\ket{1_{\boldsymbol{s}}}$ with one photon in mode ${\boldsymbol{s}}$ are involved.
Similarly, the absorption probability amplitude is
$-i g I_{{\mathrm a}, {\boldsymbol{s}} }$, where
$I_{{\mathrm a}, {\boldsymbol{s}} } =
\int d\tau \;\bra{0,a}V_I(\tau)\ket{1_{\boldsymbol{s}},b}$.
Thus, the emission and absorption probabilities are given by
  \begin{equation}
   \left\{ \begin{array}{l}
  P_{{\mathrm e}, {\boldsymbol{s}} } \\
  P_{{\mathrm a}, {\boldsymbol{s}} }
  \end{array}\right\}
  =
  g^2 \left|\int\; d\tau\; \phi_{(\pm);\boldsymbol{s}}(\mathbf{r} (\tau),t(\tau)) \, e^{i\nu\tau}\right|^2
     \label{eq:P_explicit}
     \end{equation}
   where $\phi_{(\epsilon);{\boldsymbol{s}}}$, with $\epsilon = \pm$, are functions selected from 
   $\phi_{\boldsymbol{s}}$ and $\phi_{\boldsymbol{s}}^{*}$ 
   according to the convention 
   $\phi_{(\pm);{\boldsymbol{s}}} = \phi_{\boldsymbol{s}}^{*}, \phi_{\boldsymbol{s}}$ (in that order).
   
   \subsection{Field density matrix and master equation}
   Our goal is to study the the thermal properties of the HBAR radiation field.
   In order to find the field configuration that is generated by the falling atomic cloud in the Kerr geometry, an additional ingredient is useful beyond the 
the geometry details and interactions discussed above: the density matrix.
This is needed to fully characterize the nature of the emerging field state and to compute the thermodynamic properties. The details are fully worked in the first article of this series~\cite{HBAR_part-I},
   generalizing Refs.~\cite{scully2003} and \cite{belyanin2006}; a brief summary follows next.
   The key step consists in evaluating the rate of change of the reduced density matrix 
    ($\rs{\mathcal P}$) of the field, due to the random injection of atoms. The reduced field density matrix 
%
%
is obtained via partial tracing (over the atomic degrees of freedom) from the density matrix of the composite 
system:
 $\rs{\mathcal P}  =  \mathrm{Tr_{\mathcal A}} \, \left(  \rs{{\mathcal P}{\mathcal A}} \right)$.
 At the same time, this requires enforcing an averaging procedure with respect to the atomic cloud 
 (distribution of injection times)
 in going from the density matrix of one atom to that of the whole cloud.
 The resulting coarse-grained field density matrix
 satisfies the multimode master equation~\cite{HBAR_part-I}
\begin{equation}
\begin{aligned}
 \dot{\rho}_{\rm diag}(  \boldsymbol{ \left\{  \right. } n  \boldsymbol{\left. \right\}  } )  
 =
    - 
     \sum_{j}
     &
     \left\{
     R_{{\rm e},\, j}  \big[(n_j+1) \,
   {\rho}_{\rm diag} (  \boldsymbol{ \left\{  \right. } n  \boldsymbol{\left. \right\}  } )
      - n_j \,
   {\rho}_{\rm diag} (  \boldsymbol{ \left\{  \right. } n  \boldsymbol{\left. \right\}  }_{n_j -1} )
            \big] \right.
          \\
           &
            \left.
            +  
        R_{{\rm a},\, j} \big[ n_j \,
    {\rho}_{\rm diag} (  \boldsymbol{ \left\{  \right. } n  \boldsymbol{\left. \right\}  } )
      - (n_j + 1)  \,
     {\rho}_{\rm diag} (  \boldsymbol{ \left\{  \right. } n  \boldsymbol{\left. \right\}  }_{n_j +1} )
                           \big] \right\}
    \label{eq:master_equation_final_multimode}
    \; ,
\end{aligned}
\end{equation}
which is valid under the assumption 
 that only the diagonal elements are relevant;
this is the case for {\em random injection times\/}.
In Eq.~(\ref{eq:master_equation_final_multimode}),
the emission and absorption rate coefficients are
 $R_{e, j } = \mathfrak{r} \, P_{e, j }$
and
$ R_{a, j}= \mathfrak{r} \, R_{a, j} $,
with $\mathfrak{r}$ being the atom injection rate; and
the index $j$ is shorthand for a given mode ${\boldsymbol{s}}_{j}  $,
with the single-mode quantum numbers $ {\boldsymbol{s}} $ chosen in an ordered sequence.
In addition,
the diagonal elements of the density matrix are denoted by
$  
  {\rho}_{\rm diag} (  \boldsymbol{ \left\{  \right. } n  \boldsymbol{\left. \right\}  } )
   \equiv
  {\rho}_{  n_1,n_2, \ldots  ;   n_1,n_2, \ldots  }
$,
where the notation 
$ \boldsymbol{ \left\{  \right. } n  \boldsymbol{\left. \right\}}    \equiv
\boldsymbol{ \left\{  \right.  } n_{1}, n_{2}, \ldots , n_{j } , \ldots \boldsymbol{\left. \right\}  }$ 
is used for the occupation number representation, along with
$
  \boldsymbol{ \left\{  \right. } n  \boldsymbol{\left. \right\}}_{n_j + q}  
  \equiv
 \left\{  \right.  
 n_{1}, n_{2}, \ldots , n_{j }+ q , \ldots \boldsymbol{\left. \right\}  }
 $ (with $q$ an integer-number shift).
   In Sec.~\ref{sec:conformal_steady_state},
    we will use Eq.~(\ref{eq:master_equation_final_multimode}) to establish the thermal nature of 
    the HBAR radiation field.

\section{Near-horizon physics and conformal quantum mechanics in Kerr geometry:
Field equations and geodesics}
\label{sec:nh-CQM_field-geodesics}
Two direct consequences of the gravitational field are need for the calculations of the HBAR radiation field:
the equations satisfied by the scalar field modes and the geodesic equations.

A scalar field with mass $\mu_\Phi$ 
in a generic metric $g_{\mu \nu}\,\partial_{\nu}$
satisfies the Klein-Gordon equation 
\begin{equation}
\left( \Box - \mu_\Phi^{2}  \right) \Phi \equiv \frac{1}{ \sqrt{-g} }\partial_{\mu} \left(\sqrt{-g} \,g^{\mu \nu}\,\partial_{\nu} \Phi\right)- \mu_\Phi^{2} \Phi= 0
\; .
\label{eq:Klein_Gordon_basic}
\end{equation}
 The general form of the metric leads to a differential equation for the mode functions that is a particular case of the Teukolsky equation~\cite{frolov}. Instead of tackling its general form, 
 we will mainly consider its particular near-horizon behavior, which leads to a simplified governing equation of motion for the field that highlights the physical role played by conformal symmetry. In this section, following Ref.~\cite{azizi2021}, we briefly discuss the near-horizon form of the field modes.

The field modes can be found by the following procedure.
The Kerr metric~(\ref{eq:Kerr1}) has the symmetries of independence with respect to $t$ and $\phi$ 
(stationary and axisymmetric); this implies the existence of the Killing vectors
$ \partial_t$ and $ \partial_\phi$ of Eq.~(\ref{eq:Killing-vectors}).
As a result, it can be analyzed with the separation of variables 
\begin{equation}
    \phi_{\boldsymbol{s}} (r,\Omega, t) 
    = R_{\boldsymbol{s}} (r) S_{\boldsymbol{s}}(\theta) e^{im\phi} e^{-i\omega t} 
       \; ,
          \label{eq:KG_separation}
\end{equation}
Replacing Eq.~(\ref{eq:KG_separation}) in Eq.~(\ref{eq:Klein_Gordon_basic}) for the metric~(\ref{eq:Kerr1}), the polar-coordinate angular equation is satisfied by spheroidal wave functions~\cite{frolov,spheroidal}.
We will use the regular solutions, the oblate spheroidal wave functions of the first kind
$S_{\boldsymbol{s}}(\theta)$~\cite{spheroidal},
with eigenvalues given by the separation costant $ \Lambda_{{\boldsymbol{s}}} $; 
the normalized combination $  Z_{\boldsymbol{s}}(\Omega) =(2 \pi)^{-1/2}
S_{\boldsymbol{s}}(\theta) e^{im\phi}  $ is a spheroidal harmonic.
Then, the radial function $R(r)$ becomes
\begin{equation}
 \frac{d}{d r} \left( \D  \frac{d R}{d r} \right) +
\left[\frac{(r^2+a^2)^2\omega^2 - 4 Mra m \omega + a^2 m^2}{\D} - \Lambda_{{\boldsymbol{s}}} - a^2 \omega^2 - \mu_\Phi^2 r^2
\right] R = 0
\label{eq:Kerr_Klein_Gordon_radial}
\; .
\end{equation}
In addition, we introduce a new set of coordinates 
\begin{equation}
\Tilde{t}=t  
\; , \; \; \; 
\Tilde{\phi}=\phi-\Omega_H t
\; ,
\label{eq:rotating-coords_separation_1}
\end{equation}
which define a corotating frame 
with the black hole's angular velocity $\Omega_{H}$.
At the same time, we perform a concomitant separation of variables
\begin{equation}
\phi_{\tilde{\omega} l m} (\boldsymbol{r}, t) 
= R(r) S(\theta) e^{im\tilde{\phi}}  e^{-i\tilde{\omega} \tilde{t}}
\; , \; \; \; 
\tilde{\omega}=\omega-m\Omega_H
\; ,
\label{eq:rotating-coords_separation_2}
\end{equation}
which is viewed by a locally corotating observer as a frequency shift.
Equation~(\ref{eq:rotating-coords_separation_2}) highlights the 
shifted frequency $\tilde{\omega}$ to be used for the remainder of this paper; its relevance is due
to its association with the Killing vector  
\begin{equation}
 \boldsymbol\xi \equiv \boldsymbol\xi_{(\tilde{t})} 
=  \boldsymbol\xi_{(t)} + \Omega_{H} \boldsymbol\xi_{(\phi)}
\; ,
\label{eq:Killing-vector_tilde-t}
\end{equation}
 which is timelike near the event horizon and null on the event horizon. 
 Instead, the original Killing vector $\partial_t$ is spacelike near the horizon and 
throughout the interior of a region known as the ergosphere, whose external boundary---the outer static limit---is defined as the locus where $\partial_t$ or $g_{tt}$ is null. The timelike behavior of $\xi_{(\tilde{t})}$ allows us to define positive frequency modes near the horizon with respect to this Killing vector, from the condition
 $\xi_{(\tilde{t})} \phi_{\boldsymbol{s}}  = - i \tilde{\omega} \phi_{\boldsymbol{s}}$.

The near-horizon expansion, denoted by $\stackrel{(\mathcal H)}{\sim}$,
involves the use of the hierarchy $x \equiv r-r_+ \ll r_+$.
 This can be enforced with the substitutions
\begin{eqnarray}
\D(r)  \stackrel{(\mathcal H)}{\sim} & \D'_{+}  \, x \left[ 1 + O(x) \right]
\; , \; \; \; 
\D'(r)  \stackrel{(\mathcal H)}{\sim} & \D'_{+} \left[ 1 + O(x) \right]
\; , \; \; \; 
\D''(r)  =   \D'_{+}  =2 
\, .
\end{eqnarray}
In particular, we will apply this procedure to both the field equations and the geodesics.

For the Klein-Gordon equation~(\ref{eq:Klein_Gordon_basic}), with the
expression for the Kerr metric given by Eq.~(\ref{eq:Kerr3}),
in the near-horizon region, one can directly write
\begin{align}
&
\left[
- \frac{\Sigma^2}{\rho^2 \Delta} \frac{\partial}{\partial  \tilde{t}^2}
+
\frac{\rho^2 }{\Sigma^2 \s^2 \T } \frac{\partial}{\partial  \tilde{\phi}^2}
+
\frac{1}{\rho^2} \frac{\partial}{\partial  r} \left( \D  \frac{\partial}{\partial r} \right)
+
\frac{1}{\rho^2} \frac{\partial}{\partial  \T^2}
\right]
\Phi
\\
&
\stackrel{(\mathcal H)}{\sim}
\left[
- \frac{(r^2+a^2)^2}{\rho^2 \Delta} \frac{\partial}{\partial  \tilde{t}^2}
+
\frac{1}{\rho^2} \frac{\partial}{\partial  r} \left( \D  \frac{\partial}{\partial r} \right)
\right] \Phi
\stackrel{(\mathcal H)}{\sim}
0
\; ;
\label{eq:Kerr_Klein_Gordon_conformal-with-time}
\end{align}
this is governed by the leading behavior 
$\D(r)  \stackrel{(\mathcal H)}{\sim}  \D'_{+}  \, x$, which extracts the radial-time sector of the metric.
Equation~(\ref{eq:Kerr_Klein_Gordon_conformal-with-time})
yields
\begin{equation}
\left[\frac{1}{x} \frac{d}{d  x} \left( x  \frac{d}{d x} \right) 
+ \left( \frac{ \tilde{\omega}}{2 \kappa} \right)^{2}
\frac{1}{x^2} \right] R(x)
\stackrel{(\mathcal H)}{\sim} 0 \; .
\label{eq:Kerr_Klein_Gordon_conformal-R}
\end{equation}
It is easy to verify that this near-horizon equation can also be derived from Eq.~(\ref{eq:Kerr_Klein_Gordon_radial}) with additional algebra.

Equation~(\ref{eq:Kerr_Klein_Gordon_conformal-R}) is conformally invariant and can be reduced to the
 standard form of CQM  with the Liouville transformation $R(x) \propto x^{-1/2} u(x)$;
 thus, the near-horizon reduced radial function $u(x)$ satisfies the differential equation
\begin{equation}
\frac{d^2 u(x) }{d  x^2} + \frac{ \lambda }{x^{2}} \,
\left[ 1 + \mathcal{O}(x) \right] u (x) = 0\;  ,
\label{eq:Kerr_Klein_Gordon_conformal}
\end{equation}  
where 
\begin{equation}
\lambda = \frac{1}{4} + \Theta^2\, , \hskip 4em \Theta = \frac{\tilde{\omega}}{2\kappa} 
\; .
\end{equation}
As in Ref.~\cite{HBAR_part-I},
the Hamiltonian $\mathscr{H} = p_x^2/2 + V_{\rm eff}(x)$, where $V_{\rm eff}(x) = - \lambda/x^2$ is classically scale invariant and defines
 conformal quantum mechanics (CQM)
with an enlarged SO(2,1) symmetry group
that includes 
$\mathscr{H} $,  the dilation operator $D$ and the special conformal operator $K$. 

Equation~(\ref{eq:Kerr_Klein_Gordon_conformal}) leads to a basic set of solutions given by $u(x) = x^{1/2\pm i\Theta}$. These are outgoing/ingoing wave functions that possess a 
 logarithmic-phase singular behavior associated with scale invariance.
When their time dependence is restored, these solutions give the outgoing and ingoing CQM modes, 
\begin{equation}
    \phi_{\boldsymbol{s}} (r,\Omega, t) \stackrel{(\mathcal H)}{\sim} \Phi^{\pm {\rm \scriptscriptstyle (CQM)}}_{\boldsymbol{s}} \propto 
     x^{\pm i\Theta} S_{\boldsymbol{s}}(\theta) e^{im\tilde{\phi}} e^{-i\tilde{\omega} t} 
     \label{eq:CQM_modes}
     \; ,
\end{equation}
where $\tilde{\phi}$ is the redefined corotating azimuthal coordinate for spinning black holes.
We will use these CQM modes in Sec.~\ref{sec:conformal_steady_state} to find the emission and absorption rates of the free-falling atoms.

We now turn to a brief summary of relevant results of the near-horizon geodesics in the Kerr geometry. 
These are the spacetime trajectories of the atoms in free fall---they are in locally inertial systems in the black-hole background, according to general relativity. The trajectories are required 
 for the evaluation of the photon emission and absorption probabilities in Eq.~(\ref{eq:P_explicit}),
 which we will consider in the near-horizon approximation.
The geodesic equations are derived and analyzed in Appendix~\ref{app:Kerr-geodesics}. 
Inspection of Eq.~(\ref{eq:P_explicit}) shows that the functional dependences $x^{\mu}(\tau)$ are needed. 
Basically, the relationship between $r$ and $\tau$ can be easily determined in the near-horizon limit, leading to explicit expressions of the other coordinates in terms of $x=r-r_{+}$.
In short, the leading near-horizon relationship $\tau= \tau(x)$ is
 \begin{equation}
\tau \stackrel{(\mathcal H)}{\sim} -k  x+ \mo(x^2) + \text{const} 
\\; ,
\label{eq:tau-vs-x}
\end{equation}
where $k=\hat{\rho}_+^2/c_0$;
and the relationships $t = t(x)$ and $\phi= \phi(x)$ take the forms
\begin{align}
    t  &\stackrel{(\mathcal H)}{\sim} - \frac{1}{2 \kappa} \ln x - C x +\mo(x^2) \; , \label{eq:t-vs-x} \\
    \tilde{\phi} &\stackrel{(\mathcal H)}{\sim} \alpha x +\mo(x^2)\; . 
     \label{eq:phi-vs-x}
\end{align}
where the constants $C$ and $\alpha$ are given in Appendix~\ref{app:Kerr-geodesics}. 
The derivation of the emission and absorption probabilities in the next section shows that 
neither these constants nor $k$ are explicitly needed for the final radiation results.
 
\section{HBAR in Kerr geometry:
Conformal nature and thermal characterization}
 \label{sec:conformal_steady_state}
 \subsection{Emission and absorption probabilities: conformal nature}
Emission and absorption rates of scalar photons by freely falling atoms in the Kerr geometry
can be derived from the basic Eq.~(\ref{eq:P_explicit}) with the 
 outgoing CQM modes from Eq.~(\ref{eq:CQM_modes}), with the assignments
   $\phi_{(+);{\boldsymbol{s}}} \sim
\left[ \Phi^{+ {\rm \scriptscriptstyle (CQM)}} \right]^{*}$
 and
    $\phi_{(-);{\boldsymbol{s}}} \sim
 \Phi^{+ {\rm \scriptscriptstyle (CQM)}}$; then,
   \begin{equation}
   \left\{ \begin{array}{l}
  P_{{\mathrm e}, {\boldsymbol{s}} } \\
  P_{{\mathrm a}, {\boldsymbol{s}} }
  \end{array}\right\}
\stackrel{(\mathcal H)}{\sim}\; 
g^2k^2 \left|\int_0^{x_f} \; dx\; 
x^ {\mp i\Theta} e^{ \mp i m \tilde{\phi}(x)} e^{ \pm i\tilde{\omega}t(x)} e^{i\nu\tau(x)} \right|^2
    \label{eq:P_CQM}
     \end{equation}
   where
   $k=\hat{\rho}_+^2/c_0$ and we have kept the $\tilde{\phi}$ dependence for consistency of near-horizon calculations in the presence of frame dragging. 
   In Eq.~(\ref{eq:P_CQM}), the near-horizon region is bounded, in principle,  by an approximate upper limit $x_f$.
   In particular, when $\tau$, $t$, and $\tilde{\phi}$ are replaced in the integrand of Eq.~(\ref{eq:P_CQM}) with
  Eqs.~(\ref{eq:tau-vs-x}), (\ref{eq:t-vs-x}), and (\ref{eq:phi-vs-x}), the emission rate
  takes the form
\begin{equation}
        R_{e, {\boldsymbol{s}} }   =  \mathfrak{r} \, g^2 k^2 \left|\int_0^{x_f} 
        dx \;x^{-i\tilde{\omega}/\kappa} e^{-i q x} \right|^2 \; ,
    \label{eq:near-horizon-pex}
\end{equation}
where $q=C\tilde{\omega}+k \nu + \alpha m$,
with $C$ and $\alpha$ given by Eqs.~(\ref{eq:constant-C_Kerr}) and (\ref{eq:constant-alpha}), respectively.
 
 The emission rate of Eq.~(\ref{eq:near-horizon-pex}) displays an integral that
  has been analyzed in Refs.~\cite{HBAR_part-I,camblong2020,azizi2021}
 The conclusions of that analysis are as follows.
First, the behavior of the integral is controlled by the competition of 
the two oscillatory factors $x^{-i\tilde{\omega}/\kappa}$ and $e^{-iqx}$, 
 Second, the factor $x^{-i\tilde{\omega}/\kappa}$, which arises from the properties of CQM,
oscillates wildly with a logarithmic phase as the horizon is approached, with scale invariance governed by 
 $\tilde{\omega}/\kappa = 2\Theta$, and is thus responsible for the leading value of the integral.
 Third, the upper limit $x_{f}$ can be effectively replaced by infinity as the oscillations of $e^{-iqx}$
average out to essentially zero.
 Consequently, the final expression for the emission rate becomes
 \begin{equation}
    R_{e, {\boldsymbol{s}} }  =  
     \mathfrak{r}  \, g^2 k^2
       \left|\int_0^{\infty} \; dx\;  x^{-i\tilde{\omega}/\kappa} e^{-i{q} x} \right|^2 = \frac{2\pi  \mathfrak{r} \,  g^2 \tilde{\omega}}{\kappa\,\nu^2}\;\frac{1}{e^{2\pi\tilde{\omega}/\kappa}-1}\;,
    \label{eq:R_ex_steps} 
\end{equation}
in which the approximation $\nu\gg \tilde{\omega}$ is applied.
The functional dependence of the emission rate in Eq.~(\ref{eq:R_ex_steps}) corresponds to a Planck distribution; in particular, it is insensitive to the initial conditions of the atoms in the cloud because 
it does not depend on the constants $k$, $C$, and $\alpha$.
 
 Moreover, while the outgoing CQM waves in Eq.~(\ref{eq:CQM_modes}),
$ \Phi^{+ {\rm \scriptscriptstyle (CQM)}}_{\boldsymbol{s}}
\propto x^{ i\Theta} S_{\boldsymbol{s}}(\theta) e^{im\tilde{\phi}} e^{-i\tilde{\omega} t}$,
give the Planck-distribution form of Eq.~(\ref{eq:R_ex_steps}), 
the ingoing waves $ \Phi^{+ {\rm \scriptscriptstyle (CQM)}}_{\boldsymbol{s}}$ do not contribute due to the cancellation of the logarithmic phases.
An important consequence of this property is that acceleration radiation with a Planckian distribution from a freely falling atom will exist for any generic Boulware-like state $\left| B \right\rangle$, as a result of the nonzero conformal integral in Eq.~(\ref{eq:R_ex_steps}).
In effect, all that is needed is the extraction of the outgoing part of any definition of the field modes.
For the implementation of this procedure, one simple choice of boundary conditions and vacuum is
to analyze the acceleration radiation and its associated HBAR entropy 
consists in enclosing the system with a boundary or mirror inside the speed-of-light surface (which is within the ergosphere)~\cite{rotating-Qvacuum}, thus removing the problematic superradiant modes~\cite{frolov} and defining a unique state
$\left| B_{\mathcal M} \right\rangle$, as considered in the references under~\cite{mirror-no-superradiance}.
Alternatively, as analyzed in Appendix~\ref{app:vacuum_modes}, the conventional bases that define the past and future Boulware states $\left| B^{\mp} \right\rangle$~\cite{ottewill00,menezes17}
require the selection of purely outgoing components---these are $ \phi^{\mathrm{up}}_{\omega l m} $ for $\left| B^{-} \right\rangle$ and 
$\phi^{\mathrm{out}}_{\omega l m} $ for $\left| B^{+} \right\rangle$. The precise definitions and characterizations of these states are reviewed in Appendix~\ref{app:vacuum_modes}.
For the states $\left| B^{\mp} \right\rangle$, the superradiant modes~\cite{Unruh-Starobinsky-rad}
 (for $- m \Omega_{H} < \tilde{\omega} < 0$) need to be handled separately. But, most importantly, the outgoing components can be identified as in Eq.~(\ref{eq:CQM-vs-global-modes}) for
 $\left| B^{\pm} \right\rangle$.
 In short, all of the possible definitions of a Boulware-like state---including in particular
 $\left| B_{\mathcal M} \right\rangle$ and  $\left| B^{\pm} \right\rangle$---directly yield 
 the Planck-distribution form of Eq.~(\ref{eq:R_ex_steps}) through the extraction of the
 outgoing components, with any remaining ingoing components not contributing to the radiation.

Finally, the absorption rate $R_{a, {\boldsymbol{s}} }$ can be computed in a manner similar to the derivation leading to the emission rate $R_{e, {\boldsymbol{s}} }$ of Eq.~(\ref{eq:R_ex_steps}).  As it turns out, $R_{a, {\boldsymbol{s}} }$ is more directly derived  from $R_{e, {\boldsymbol{s}} }$ via the replacements  $\tilde{\omega} \rightarrow -\tilde{\omega}$, and $ m \rightarrow -  m$;   this is, from first principles, due to the functional forms of the rates in Eq.~(\ref{eq:P_CQM}).   Thus,
\begin{equation}
   R_{a, {\boldsymbol{s}} }
     = \frac{2\pi r g^2 \tilde{\omega}}{\kappa\,\nu^2}\;\frac{1}{1-e^{-2\pi\tilde{\omega}/\kappa}}
     =
 e^{2 \pi \tilde{\omega}/\kappa }  \,      R_{e, {\boldsymbol{s}} } 
     \label{eq:R_ab}
     \;  ,
\end{equation}
which has remarkable consequences for the question of possible thermal behavior,
 as will be shown next.

 \subsection{Thermal behavior}
 We will now analyze in detail the properties of the HBAR radiation field and its possible thermal nature.
 This involves two levels. First, we will reexamine the 
 emission and absorption rates to evaluate a candidate Boltzmann 
 factor and its relationship to the Hawking temperature.
 Second, we will conduct a more thorough analysis through the reduced field density matrix.
 
 The Planck distribution displayed by the emission rate in Eq.~(\ref{eq:R_ex_steps}) suggests 
 thermal behavior---or at least, it has a universal form compatible with thermality for all modes of the radiation field. This property can be further tested with the ratio emission over absorption, which, from
 Eq.~(\ref{eq:R_ab}), is equal to
\begin{equation}
    \frac{R_{e, {\boldsymbol{s}} } }{R_{a, {\boldsymbol{s}} } } = e^{-2\pi\tilde{\omega}/\kappa}\,. \label{eq:ratio_em_ab}
\end{equation}
This ratio has the form of a thermal state with detailed-balance Boltzmann factor
\begin{equation}
    \frac{R_{e, {\boldsymbol{s}} } }{R_{a, {\boldsymbol{s}} } } = e^{-\beta \tilde{\omega}}
    \; 
    \label{eq:ratio_em_ab_Boltzmann}
\end{equation}
(in natural units where the Boltzmann constant is $k_{B} =1$), with temperature
 \begin{equation}
T= \beta^{-1}  = \frac{\kappa}{2\pi} \equiv \beta_{H}^{-1} = T_H
\; .
\label{eq:temperature=Hawking}
\end{equation}
 The value of the effective temperature defined by this procedure in 
Eq.~(\ref{eq:temperature=Hawking}) is the same as the Hawking temperature of the black hole---and
it agrees with the form of the Planck distribution of Eq.~(\ref{eq:R_ex_steps}). 

It should be noted that the ``temperature'' in the primary thermal properties of
Eqs.~(\ref{eq:ratio_em_ab_Boltzmann}) and (\ref{eq:temperature=Hawking}) does not appear to be 
merely an effective phenomenological parameter,
but it is good candidate for a thermodynamic temperature in detailed-balance relations. 
This is due to the fact that it is a universal temperature defined for all modes in 
Eq.~(\ref{eq:ratio_em_ab_Boltzmann}), 
which is identical to the thermodynamic temperature of the black hole.
In addition, the condition displayed in Eq.~(\ref{eq:ratio_em_ab_Boltzmann})
is a direct consequence of the equations of near-horizon CQM, and can be traced to the same logarithmic form 
of the phase in all of its modes.

The thermal nature of the state of the HBAR radiation field, under the conditions defined via CQM in
Eqs.~(\ref{eq:ratio_em_ab_Boltzmann}) and (\ref{eq:temperature=Hawking}),
can be fully characterized using the master equation for the field density matrix. 
  For a cloud of freely falling atoms, with random injection times,
  the density matrix has a diagonal form and satisfies the master equation~(\ref{eq:master_equation_final_multimode}).
 This is the procedure fully worked out in Ref.~\cite{HBAR_part-I}, and which we now summarize and generalize for the Kerr geometry. In fact, the general properties of the density matrix in this approach
 are essentially geometry-independent and apply equally well to the
 generalized Schwarzschild and Kerr geometries.

In Eq.~(\ref{eq:master_equation_final_multimode}),
the vanishing of the time derivative  defines
the steady-state density matrix 
$  {{\rho}}^{\mathrm (SS)}_{\rm diag}(  \boldsymbol{ \left\{  \right. } n  \boldsymbol{\left. \right\}  } )$.
 Finding the steady-state solution involves an effective procedure that consists of the following steps.
 First, one starts by finding the single-mode steady state, which satisfies
  \begin{equation}
\left. {\rho}_{n,n}^{\mathrm (SS)} \right|_{\mathrm{single-mode}}
 = 
  \left[ 1-\left( \frac{ R_{e, {\boldsymbol{s}} } }{ R_{a, {\boldsymbol{s}}} } \right)  \right]
  \,
 \left( {R_{e, {\boldsymbol{s}} } }/{ R_{a, {\boldsymbol{s}} }} \right)^n
 =
 \frac{1}{Z_{j}}
    e^{-n_{j} \beta \tilde{\omega}_{j}}
 \; ,
     \label{eq:steady_state_single-mode_thermal}
 \end{equation}
 with $Z_{j} = \left[ 1- e^{- \beta \tilde{\omega}_{j}}\right]^{-1}$.
 Second, the factorization
\begin{equation}
 {\rho}_{\rm diag}^{\mathrm (SS)}(  \boldsymbol{ \left\{  \right. } n  \boldsymbol{\left. \right\}  } )
 =
 \prod_{j} {\rho}_{n_{j},n_{j}}^{\mathrm (SS)} 
 \; 
 \label{eq:SS-density-matrix_factorization}
\end{equation}
 can be proposed because the multimode density matrix is expected to be
 composed of independent single-mode pieces under the given injection-averaging condition.
  Third, the fact that this is the correct state can be verified by direct substitution in
  Eq.~(\ref{eq:master_equation_final_multimode}).
  Fourth, the effective temperature $T =\beta^{-1}$ can be enforced from the Boltzmann factor.
  As a result,
 \begin{equation}
    {{\rho}}^{\mathrm (SS)}_{\rm diag}(  \boldsymbol{ \left\{  \right. } n  \boldsymbol{\left. \right\}  } )
= N \, \prod_{j}  \left(  \frac{R_{e, j } }{ R_{a, j} } \right)^{n_{j}} 
  = \frac{1}{Z}
   \prod_{j}  
    e^{-n_{j} \beta \tilde{\omega}_{j}}
    \; ,
    \label{eq:steady_state_multimode_combined}
\end{equation}
where $Z= N^{-1}= 
\prod_{j} Z_{j}
= \prod_{j} 
\left[ 1- e^{- \beta \tilde{\omega}_{j}}\right]^{-1}$ is the partition function.
This is indeed a thermal distribution at the Hawking temperature, according to
Eqs.~(\ref{eq:ratio_em_ab_Boltzmann}) and (\ref{eq:temperature=Hawking});
in particular, it yields the steady-state average occupation numbers
$ \left\langle n_{j} \right\rangle \! ^{\! \! ^{\mathrm (SS)}}  = \left( e^{\beta \tilde{\omega}_{j}} -1 \right)^{-1}$.

Two critical points should be highlighted:
(i) the appearance of the shifted frequency $\tilde{\omega}$ 
in all ensuing expressions describing the thermal behavior;
(ii) the fact that the result is critically dependent on the primary thermal properties of
Eqs.~(\ref{eq:ratio_em_ab_Boltzmann}) and (\ref{eq:temperature=Hawking}): 
the Boltzmann factor and the Hawking temperature, where both emerge from near-horizon CQM for
all field modes. 
The thermal state  of Eq.~(\ref{eq:steady_state_multimode_combined})
confirms and extends the validity of the steady-state analysis and HBAR properties, including
the entropy flux calculations,
of Refs.~\cite{scully2018, HBAR_part-I} for Kerr black holes. 
 Such surprising results show close parallels with the thermodynamics of the black hole itself~\cite{scully2018}. We now turn to an analysis of this all-important problem, with a thorough treatment of HBAR thermodynamics.

\section{HBAR thermodynamics: 
Entropy and HBAR-black-hole correspondence} 
\label{sec:HBAR-thermo}

\subsection{From von Neumann to the thermodynamic entropy}
\label{sec:entropy_vN-to-thermo}

In Sec.~\ref{sec:conformal_steady_state},
we have derived the thermal steady-state density matrix directly from near-horizon CQM, and generalized it to rotating black holes with arbitrary initial conditions for the freely falling atoms. With these robust results, we can now proceed to find the entropy rate of change or flux $ \dot{S}_{\mathcal P}$ due to the generation of acceleration radiation or field quanta (``photons''). This is the horizon brightened acceleration radiation (HBAR) entropy proposed in Ref.~\cite{scully2018}, whose insightful procedure we will follow and generalize in this section.

In general, for any quantum system, starting from the von Neumann entropy, 
$S = - \mathrm{Tr} \left[  {\rho} \ln {\rho} \right] $,
its rate of change is simply given by $\dot{S} = - \mathrm{Tr} \left[  \dot{{\rho}} \ln {{\rho}} \right] $.
For the radiation field, the corresponding trace can be evaluated as
\begin{equation}
 \dot{S}_{\mathcal P} =  -  \sum_{\boldsymbol{ \left\{  \right. } n  \boldsymbol{\left. \right\}  } } 
 \dot{{\rho}}_{\rm diag}(  \boldsymbol{ \left\{  \right. } n  \boldsymbol{\left. \right\}  } )
 \ln  \left[   {\rho}_{\rm diag}(  \boldsymbol{ \left\{  \right. } n  \boldsymbol{\left. \right\}  } ) \right]
 =
- \sum_{  n_1,n_2, \ldots }
  \dot{\rho}_{ n_1,n_2, \ldots ;   n_1,n_2, \ldots }
 \ln   {\rho}_{ n_1,n_2, \ldots ;   n_1,n_2, \ldots } 
    \; ,
    \label{eq:entropy_rate-change}
\end{equation}
in which we are considering the diagonal sum over all the states
${\boldsymbol{ \left\{  \right. } n  \boldsymbol{\left. \right\}  } }$ associated with all the field modes,
as in Sec.~\ref{sec:conformal_steady_state}.
Near the steady-state configuration, the density matrix in the logarithm of Eq.~(\ref{eq:entropy_rate-change}) can be approximated to leading order with ${\rho}^{\mathrm (SS)}_{\rm diag}$ from
Eq.~(\ref{eq:steady_state_multimode_combined}). Thus, the entropy flux of Eq.~(\ref{eq:entropy_rate-change}) becomes
\begin{equation}
    \dot{S}_{\mathcal P} \approx 
 - \sum_{\boldsymbol{ \left\{  \right. } n  \boldsymbol{\left. \right\}  } } 
  \dot{{\rho}}_{\rm diag}(  \boldsymbol{ \left\{  \right. } n  \boldsymbol{\left. \right\}  } ) 
 \ln  \left[   {\rho}^{\mathrm (SS)}_{\rm diag}(  \boldsymbol{ \left\{  \right. } n  \boldsymbol{\left. \right\}  } ) \right]
 =
 - \sum_{j}  \sum_{\boldsymbol{ \left\{  \right. } n  \boldsymbol{\left. \right\}  } } 
  \dot{{\rho}}_{\rm diag}(  \boldsymbol{ \left\{  \right. } n  \boldsymbol{\left. \right\}  } ) 
\ln  {\rho}_{n_{j},n_{j}}^{\mathrm (SS)}
    \; ,
    \label{eq:entropy_rate-change_SS}
\end{equation}
as follows from the factorization~(\ref{eq:SS-density-matrix_factorization}) and by reversing the order of the sums. As a reminder, the resulting summation with respect to $j$ is a shorthand for the sum over all the field-mode numbers ${\boldsymbol{s}} =  \{\tilde{\omega},l,m\}$, i.e., the field frequencies $\tilde{\omega}$ in addition to the angular quantum numbers.
It should be noted that
it is essential to account for all the available frequencies in appropriate sums that define the full-fledged thermodynamic behavior.
The replacement of the thermal steady-state density matrix, e.g.,  Eq.~(\ref{eq:steady_state_single-mode_thermal}), in Eq.~(\ref{eq:entropy_rate-change_SS}) implies that
\begin{equation}
    \dot{S}_{\mathcal P} \approx 
 \sum_{j}  \sum_{\boldsymbol{ \left\{  \right. } n  \boldsymbol{\left. \right\}  } } 
  \dot{{\rho}}_{\rm diag}(  \boldsymbol{ \left\{  \right. } n  \boldsymbol{\left. \right\}  } ) 
  \left[
  n_{j} \beta \tilde{\omega}_{j} - \ln (1-e^{- \beta \tilde{\omega}_{j}}) 
  \right]
    \; .
  \label{eq:entropy_rate-change_SS_2}
\end{equation}
Moreover, Eq.~(\ref{eq:entropy_rate-change_SS_2})  can be interpreted by enforcing two
conditions:
the trace normalization
$ 
\mathrm{Tr} \left[  
{\rho} \right] = 1$ and the dynamic generalization of the occupation-number averages
\begin{equation}
  \left\langle n_{j} \right\rangle \equiv 
\sum_{
 \boldsymbol{ \left\{  \right. } n  \boldsymbol{\left. \right\} }
}
n_{j} \,   {{\rho}}_{\rm diag}(  \boldsymbol{ \left\{  \right. } n  \boldsymbol{\left. \right\}  } ) 
    \; ,
    \label{eq:average-occupation_dynamical}
\end{equation}
where these quantities $\left\langle n_{j} \right\rangle \equiv  \left\langle n_{ \boldsymbol{s} } \right\rangle $
are defined for each set of field-mode numbers
${\boldsymbol{s}} =  \{\tilde{\omega},l,m\}$. It should be further stressed that
Eq.~(\ref{eq:average-occupation_dynamical})
is a generalization of the steady-state average occupation numbers
 $ \left\langle n_{j} \right\rangle \! ^{\! \! ^{\mathrm (SS)}}  = \left( e^{\beta \tilde{\omega}_{j}} -1 \right)^{-1}$; as such, it is no longer given exactly by the Planck distribution, in such a way that it can generate a nonzero flux through $ \dot{\, \, \left\langle n_{ \boldsymbol{s} } \right\rangle } \neq 0$.
Moreover, the second-term in Eq.~(\ref{eq:entropy_rate-change_SS_2})  vanishes, to the same order of approximation, due to the constancy of trace normalization; and the first term yields the entropy flux
\begin{equation}
    \dot{S}_{\mathcal P} \approx \beta_{H}
 \sum_{{\boldsymbol{s}} = \{\tilde{\omega},l,m\} } 
  \! \!   \! \! 
   \dot{\, \, \left\langle n_{ \boldsymbol{s} } \right\rangle }  \, \tilde{\omega} 
   =
   \frac{2\pi}{\kappa}
 \sum_{{\boldsymbol{s}} = \{\tilde{\omega},l,m\} } 
  \! \!   \! \! 
   \dot{\, \, \left\langle n_{ \boldsymbol{s} } \right\rangle }  \, \tilde{\omega} 
    \; .
    \label{eq:Sp_1}
\end{equation}
where the unique Hawking temperature, Eq.~(\ref{eq:temperature=Hawking}), is explicitly used as a last step.

In Eq.~(\ref{eq:Sp_1}), 
the product $ \dot{\left\langle n_{ \boldsymbol{s} } \right\rangle }  \, \tilde{\omega} $ is the portion of the energy flux carried away by the photons of the acceleration radiation, at a given frequency, and measured in the corotating frame. For a single photon, the corotating energy
${\tilde{\omega}} = \omega - \Omega_{H} m $ involves the energy $\omega$ (measured by an asymptotic observer) and the axial component $m$ of the angular momentum (along the black hole's rotational axis). Then, the total corotating energy flux is
\begin{equation}
\dot{\tilde{E}}_{\mathcal P}
=
 \sum_{{\boldsymbol{s}} = \{\tilde{\omega},l,m\} } 
  \! \!   \! \!  
   \dot{\, \, \left\langle n_{ \boldsymbol{s} } \right\rangle }      \, \tilde{\omega} 
   =
    \sum_{{\boldsymbol{s}} = \{\tilde{\omega},l,m\} } 
  \! \!   \! \!   
   \dot{\, \, \left\langle n_{ \boldsymbol{s} } \right\rangle }
\,  \bigl( \omega - \Omega_{H} m \bigr)
= \dot{E}_{\mathcal P} - \Omega_{H} \dot{J}_{{\mathcal P},z} 
\; ,
 \end{equation} 
 which involves a combination of the  change in the total energy $E_{\mathcal P} $ and axial angular momentum $J_{{\mathcal P},z} $ of the photons. 
Therefore, the HBAR von Neumann entropy flux becomes
\begin{equation}
    \dot{S}_{\mathcal P} = \beta_{H}
    \bigl( \dot{E}_{\mathcal P} -\Omega_{H} \dot{J}_{{\mathcal P},z}   \bigr)
    \; ,
    \label{eq:Sp_2}
\end{equation}
or $ \dot{S}_{\mathcal P} = \beta_{H} \dot{\tilde{E}}_{\mathcal P}$,
which can be restated in the form of thermodynamic changes
\begin{equation}
   \delta {S}_{\mathcal P} = \beta_{H} 
  \bigl(  \delta {E}_{\mathcal P} -\Omega_{H} \, \delta {J}_{{\mathcal P},z}  \bigr)
   \equiv
   \delta {S}_{\mathcal P}^{\, \rm (th)}
    \; ,
    \label{eq:Sp_3}
\end{equation}
where
${S}_{\mathcal P}^{\, \rm (th)}$ is the thermodynamic entropy.
 In other words, under near-equilibrium conditions, where the steady state is a good first-order approximation, the changes in {\it the HBAR von Neumann and thermodynamic entropies of the radiation field coincide\/},
with the functional changes of Eq.~(\ref{eq:Sp_3}).
More generally, going back to the von Neumann entropy, 
$S = - \mathrm{Tr} \left[  {\rho} \ln {\rho} \right] $, 
we can repeat the steps above to calculate ${S}_{\mathcal P}$ and directly verify this result. In effect, replacing 
$\dot{{\rho}}_{\rm diag}(  \boldsymbol{ \left\{  \right. } n  \boldsymbol{\left. \right\}  } ) $
by
${{\rho}}_{\rm diag}(  \boldsymbol{ \left\{  \right. } n  \boldsymbol{\left. \right\}  } ) $
in Eqs.~(\ref{eq:entropy_rate-change})--(\ref{eq:entropy_rate-change_SS_2}), these steps lead to the modified analogue of Eq.~(\ref{eq:Sp_1}), which is $S_{\mathcal P} = \beta (E_{\mathcal P} -F_{\mathcal P})$; here, $F_{\mathcal P}$ is the Helmholtz free energy, such that
$\beta F_{\mathcal P}=
- \ln Z = \sum_{j}\ln \left( 1 - e^{-\beta \tilde{\omega}_{j} } \right)
$, with a partition function $Z$ defined in Eq.~(\ref{eq:steady_state_multimode_combined}).
The agreement with Eq.~(\ref{eq:Sp_1}) is due to the fact that the changes of this thermodynamic entropy with fixed temperature reduce to Eq.~(\ref{eq:Sp_3}): the black hole acts as a temperature reservoir where the Hawking temperature is fixed by its characteristic parameters.

\subsection{HBAR-black-hole thermodynamic correspondence}
\label{sec:HBAR-BH_correspondence}

The entropy flux~(\ref{eq:Sp_2}) reinforces the existence of {\it universal thermodynamic relations\/} that are intrinsic to the black hole but have other manifestations through relevant probes. In other words, there are other 
aspects of black hole thermodynamics that extend the standard results and 
can be revealed by accessing the near-horizon physics. In particular, the vacuum states are probes that can generate different forms of radiation. By selecting a Boulware-like state, the existence of HBAR radiation and its associated entropy are revealed. 
Moreover, the thermodynamic changes of the Bekenstein-Hawking black hole entropy,
\begin{equation}
\delta S_{\mathrm{BH}}
= \beta_{H} \bigl( \delta M -  \Omega_H \delta J \bigr)
\; ,
  \label{eq:BH-entropy-changes}
\end{equation}
in terms of the mass $M$ and angular momentum $J$ of the black hole, display a striking formal similarity with the HBAR entropy flux of Eqs.~(\ref{eq:Sp_2}) and (\ref{eq:Sp_3}): 
they are analog relations with corresponding entropy, energy, and angular momentum variables; and they are subject to the Hawking temperature $T_{H}=\beta_{H}^{-1}$
of Eq.~(\ref{eq:temperature=Hawking}). Thus, there exists a thermodynamic correspondence
\begin{equation}
\bigl( 
{S}_{\mathcal P} , {E}_{\mathcal P} ,  {J}_{{\mathcal P},z} 
\bigr) 
\xlongleftrightarrow{\beta = \beta_{H}}
\bigl( 
S_{\mathrm{BH}} , M , J 
\bigr) 
\; \; \;
\; .
\label{eq:HBAR-BH-correspondence}
\end{equation}
In addition, these analog relations can be further extended to charged black holes (Kerr-Newman geometry), with an additional charge variable. 
Incidentally, as in Eq.~(\ref{eq:Sp_3}), the quantity $\delta \tilde{M} = \delta M -  \Omega_H \delta J $ is the black hole energy in the corotating frame. 
The correspondence~(\ref{eq:HBAR-BH-correspondence}) is not coincidental because of two key properties: 
(i) both systems are described by the first law of thermodynamics in terms of just energy 
(mass), angular momentum, and any other relevant degrees of freedom consistent with no-hair theorems~\cite{frolov};
(ii) they share the {\it same unique Hawking temperature\/}.
In essence, in thermal equilibrium, the quantum field macroscopically mimics the black hole degrees of freedom; and the black hole generates a near-horizon background governed by CQM and encoding this characteristic temperature [cf.\ arguments leading to Eq.~(\ref{eq:temperature=Hawking})].
Therefore, both {\it the HBAR and the black hole entropies have a common origin via near-horizon conformal symmetry; and the field satisfies 
thermodynamic relations formally identical to black hole thermodynamics.\/}
We will refer to this set of properties as the HBAR-black-hole thermodynamic correspondence.

\subsection{Area-entropy-flux relation and radiation correspondence}
\label{sec:area-entropy_relation}

There is an additional intriguing consequence of the thermodynamic correspondence. As it was found by Hawking~\cite{hawking74,hawking75}, the assignment 
$T_{H}=\beta_{H}^{-1}=\kappa/2 \pi$ of
 Eq.~(\ref{eq:temperature=Hawking})
fixes the values of the relationship of the entropy to other thermodynamic variables in Eq.~(\ref{eq:BH-entropy-changes}), thus determining the correct proportionality constant in the Bekenstein-Hawking entropy $ S_{\mathrm{BH}} = A/4$, so that
 \begin{equation}
\delta S_{\mathrm{BH}} = \frac{1}{4}   \delta A 
\; .
  \label{eq:BH-entropy-area-changes}
\end{equation}
This can be verified directly for a Kerr black hole, with area $A= 4 \pi (r_{+}^2 + a^2)$, 
which implies the change
$  \delta A  = 8 \pi  \delta \tilde{M}/\kappa$,
where $\kappa = \Delta_{+}'/2(r_{+}^2 + a^2)$; from the Hawking temperature~(\ref{eq:temperature=Hawking}), 
its change $\delta A$ is~\cite{frolov}
\begin{equation}
  \delta A  = 4 \beta_{H} \left( \delta M -  \Omega_H \delta J\right)
\; .
  \label{eq:BH-area-changes}
\end{equation}
In other words, once the temperature is fixed, there is a unique entropy-area relation.
Then, the HBAR-black-hole thermodynamic correspondence and the Bekenstein-Hawking entropy-area relation of Eq.~(\ref{eq:BH-entropy-area-changes}) suggest the existence of an analog entropy-area relation for the HBAR entropy, 
\begin{equation}
    \dot{S}_{\mathcal P} = \frac{1}{4} \bigl| \dot{A}_{\mathcal P} \bigr|
     \label{eq:HBAR_final}
     \; ,
\end{equation}
where $\bigl| \dot{A}_{\mathcal P} \bigr|$ is the magnitude of the
change in the area of the event horizon due to the emission of acceleration radiation.
Further justification of the HBAR area-entropy-flux relation~(\ref{eq:HBAR_final}) can be outlined as follows.
From Eq.~(\ref{eq:BH-area-changes}), the magnitude of the black hole area change is
$|\dot{A}|
=
 4 \beta_{H}
 |\dot{\tilde{M}}|
 $,
 where $\delta \tilde{M}
=  \delta M -  \Omega_H \delta J$.
If this change $ |\dot{\tilde{M}}|$ is due to photon emission in the amount
$| \dot{\tilde{M}}| = \dot{\tilde{E}}_{\mathcal P} $, then the black hole area changes by the following specific amount due to acceleration radiation, and according to Eq.~(\ref{eq:Sp_3}),   
\begin{equation}
 |\dot{A}_{\mathcal P}|
= 4 \beta_{H}
 \dot{\tilde{E}}_{\mathcal P}
 = 4  \dot{S}_{\mathcal P}
 \Longrightarrow 
  \dot{S}_{\mathcal P} = \frac{1}{4} \bigl| \dot{A}_{\mathcal P} \bigr|
     \label{eq:HBAR_final_derivation}
     \; ,
\end{equation}
which is the anticipated result.
The area-entropy-flux relation~(\ref{eq:HBAR_final}) is a surprising and convenient rule of thumb, with a geometric interpretation that restates the central result of this paper: the thermodynamic HBAR entropy property of Eq.~(\ref{eq:Sp_3}), which is mandated by the von Neumann entropy of the radiation field and
implies an HBAR-black-hole thermodynamic correspondence. Most importantly, we have established the robust form of these results under fairly general conditions (black hole geometry and initial conditions of the falling atomic cloud). However, some aspects of the area-entropy-flux relation~(\ref{eq:HBAR_final}) need to be qualified, as discussed below.

First, the HBAR radiation does not involve the photons and black hole alone, but it is mediated by the interaction of the field with the atoms. When the atoms are accounted for in the relevant equations of the radiation generation process, the
energy and angular momentum of the black hole are transferred to the radiation field, but subject to the following conservation laws,
\begin{subequations}
    \begin{equation}
        \delta M +  \delta E_{\mathcal P} + \delta E_{\mathcal A} = 0
        \; , 
        \label{eq:energy-conservation}
    \end{equation}
    \begin{equation}
       \delta J + \delta J_{{\mathcal P},z}  + \delta J_{{\mathcal A},z} = 0 
        \; , \label{eq:ang-momentum-conservation}
    \end{equation}
\end{subequations}
where $E_{\mathcal A} $ and $J_{{\mathcal A},z}$ are the energy and angular momentum of the atom, respectively. Equations~(\ref{eq:energy-conservation}) and (\ref{eq:ang-momentum-conservation}) can be combined into an energy conservation statement in the corotating frame,
    \begin{equation}
      ( \delta M -  \Omega_H \delta J )
      +   \delta \tilde{E}_{\mathcal P} 
        + \delta \tilde{E}_{\mathcal A} = 0 
        \; ,
         \label{eq:conservation}
    \end{equation}
 where  $\delta \tilde{E}_{\mathcal P} $ and $\delta \tilde{E}_{\mathcal A} $  are the field and atom corotating energy changes, respectively.
 Then, going back to Eq.~(\ref{eq:BH-area-changes}), the black hole area change is
$\dot{A} = 4 \beta_{H} \dot{\tilde{M}} $,
and Eq.~(\ref{eq:conservation}) leads to
$ \dot{A}  = - 4 \beta_{H} \bigl( \dot{\tilde{E}}_{\mathcal P}  +  \dot{\tilde{E}}_{\mathcal A} \bigr)$,
          which can be interpreted as giving two distinct contributions to the black hole area changes
        \begin{equation}
 \dot{A} =  \dot{A}_{\mathcal P}
 +  \dot{A}_{\mathcal A}
   \; ,
    \label{eq:area-changes}
\end{equation}
with one area change associated with the radiation field as in Eq.~(\ref{eq:HBAR_final}), i.e.,
 \begin{equation}
 \dot{A}_{\mathcal P} = - 4 \beta_{H} \dot{\tilde{E}}_{\mathcal P} = - 4 S_{\mathcal P}
     \; ,
    \label{eq:area-changes_photon}
\end{equation}
 but also with an extra contribution associated with the atoms,
  \begin{equation}
 \dot{A}_{\mathcal A} = - 4 \beta_{H} \dot{\tilde{E}}_{\mathcal A}
     \; ,
    \label{eq:area-changes_atom}
\end{equation}
 Therefore, when stating the HBAR area-entropy-flux relation~(\ref{eq:HBAR_final}), the area change 
 $ \dot{A}_{\mathcal P}$ under consideration is only a fraction of the total change in the black hole area.
 
Second, in Eqs.~(\ref{eq:area-changes})--(\ref{eq:area-changes_atom}), one should be careful with the signs---this is the reason that we need to use an absolute value in Eq.~(\ref{eq:HBAR_final}).
In particular, there is indeed a sign reversal for the radiation field: as it carries away positive corotating energy, this corresponds to a decrease in the area of the black hole (and increase in the HBAR entropy), according to Eq.~(\ref{eq:area-changes_photon}). This paradoxical situation is similar to the well-known corresponding statement for Hawking radiation. However, such statements do not contradict the generalized second law of thermodynamics (GSL), which refers to the sum of the entropies and not to the entropy of the black hole or of the radiation field alone; 
thus, $\delta A_{\mathcal P} \propto - S_{\mathcal P} <0$ is allowed. It should be noted, however, that the atoms carry corotating energy and fall into the black hole with $\delta \tilde{E}_{\mathcal A} <0$, thus yielding an area increase $\delta A_{\mathcal A}>0$, according to Eq.~(\ref{eq:area-changes_atom}). Typical nonrelativistic atoms contribute, in magnitude, significantly more than the radiation, so that the overall area of the black hole does increase---this net result is different from the corresponding case for Hawking radiation. 
 
Incidentally, if the atoms are injected sufficiently close to the event horizon, one could consider an effective description in which they are merely mediators for the radiation process, in such a way that, if they could be counted as part of the black hole for bookkeeping purposes. This is consistent with the dominant role played by near-horizon CQM and the fact that the horizon can be regarded as a thick membrane or stretched horizon~\cite{membrane,susskind93}, and as seen from the outside, the atoms appear to accumulate therein. In a very real sense, the black hole is their inevitable destiny. Then, one could interpret that there is a decrease in the black-hole area from the radiation emission alone.
  
 Third, in Eq.~(\ref{eq:conservation}), the black hole and radiation corotating energies can be directly related to the corresponding entropies via Eqs.~(\ref{eq:Sp_3}) and (\ref{eq:BH-entropy-changes}).
As a result, the generalized second law of thermodynamics,
$\delta S_{\mathrm total} =
\delta {S}_{\mathrm{BH}}
+
\delta {S}_{\mathcal A}
+
\delta {S}_{\mathcal P}
 \geq 0$, implies the inequality 
 $\delta S_{\mathcal A} \geq \beta_{H}
 \bigl(  \delta {E}_{\mathcal A} -\Omega_{H} \, \delta {J}_{{\mathcal A},z}  \bigr)
$, 
but no further predictions are possible without accounting for additional information about the atoms.

In short, our detailed analysis of the HBAR entropy highlights the thermodynamic entropy property of Eq.~(\ref{eq:Sp_3}), which similarly states the equivalence of the von Neumann and thermodynamic entropies of the acceleration radiation. In addition, the entropy of the radiated photon field
can be written in terms of the change in the area of the event horizon via the HBAR area-entropy-flux relation~(\ref{eq:HBAR_final}) as part of a larger set of relations within an HBAR-black-hole thermodynamic correspondence. In particular, the area-entropy-flux formula~(\ref{eq:HBAR_final})
is structurally identical to the Bekenstein-Hawking (BH) entropy, with the correct prefactor $1/4$. However, this entropy of the radiation field is due to the photon generation by the freely falling atoms---as such, it is apparently different from the intrinsic BH entropy, which is solely due to the existence of an event horizon as part of the geometry of a black hole. 

Similarly, the HBAR field is not the same as Hawking radiation, even though they share many formally identical qualities. The former requires the presence of an atomic cloud as a mediator for the radiation process, while the latter is intrinsic to the black hole. 
As a result of the mediation process and the 
conservation laws~(\ref{eq:energy-conservation})--(\ref{eq:conservation}), 
the HBAR radiation field is maintained only inasmuch as atoms are falling into the black hole, and is limited by the size of the cloud---but, of course, one can imagine a steady-state random injection that keeps the process going on indefinitely.
As time goes by, for the HBAR process, the black hole mass increases due to the dominant contribution of the falling atoms, while for the Hawking effect, 
the black hole mass decreases in the well-known evaporation scenario.
Despite these differences associated with the concomitant HBAR mediation by the atoms, 
for the random-injection conditions considered in this paper, both forms of radiation, as seen far from the black hole, have identical properties. Specifically,
they are both thermal and characterized by the same (Hawking) thermodynamic temperature 
$T_{H}= \kappa/2\pi$.
Therefore, just as for the other thermodynamic attributes, there is an HBAR-black-hole correspondence that extends beyond Eq.~(\ref{eq:HBAR-BH-correspondence})
to also include the mapping
\begin{equation}
\text{(HBAR field)}
\xlongleftrightarrow{\beta = \beta_{H}}
\text{(Hawking radiation)}
\; \; \;
\; .
\label{eq:HBAR-BH-correspondence_radiation}
\end{equation}

The area law~(\ref{eq:HBAR_final}) and the correspondence defined 
by Eqs.~(\ref{eq:HBAR-BH-correspondence}) and (\ref{eq:HBAR-BH-correspondence_radiation}) 
are suggestive of even deeper connections between the HBAR entropy and radiation field
on the one hand, 
 and the BH entropy and Hawking radiation on the other hand. This is an intriguing prospect that will be explored elsewhere.

\section{Conclusions}
 \label{sec:conclusion}
In this article, we have shown that the von Neumann HBAR entropy flux (time rate of change of the entropy) due to the acceleration radiation of a cloud of randomly injected atoms falling into a black hole, in a Boulware-like vacuum, takes a functional form that is in agreement with a specific form of thermodynamic entropy. This peculiar HBAR entropy is defined with degrees of freedom mimicking those of the black hole itself, and with  a temperature equal to the Hawking temperature. As a consequence, it generates a powerful HBAR-black-hole thermodynamic correspondence, which also includes the suggestive area-entropy-flux relation $\dot{S}_{\mathcal P} \propto  \bigl| \dot{A}_{\mathcal P} \bigr|$. This intriguing result has a striking formal similarity with the intrinsic Bekenstein-Hawking entropy of the black hole, including the proportionality constant $1/4$. The origin of the HBAR entropy has been elucidated with the thermal steady-state field density matrix of the acceleration radiation by the randomly injected atoms into the Boulware-like vacuum state. Moreover, the thermal property of the field density matrix is completely governed by the conformal near-horizon physics of the black hole. This is due to the fact that the field modes near the horizon are described by the CQM Hamiltonian; the scale symmetry of these modes is responsible for the thermality of the HBAR field, which involves both the Hawking temperature and the HBAR-black-hole thermodynamic correspondence, including the correct prefactor in the entropy formula. 

The near-horizon treatment is also useful as a practical analytical tool since it provides an asymptotically exact approximation (in the usual asymptotic WKB sense), which yields a closed-form solution for the emission and absorption probabilities that define the effective temperature. While we have illustrated its use for the specifics of acceleration radiation and HBAR entropy, its scope and computational effectiveness are not limited to these applications.

In conclusion, this paper establishes both the nature of the acceleration radiation and the HBAR-black-hole thermodynamic correspondence for a fairly large class of black hole solutions and arbitrary initial conditions of the atomic cloud. Furthermore, in this work, we have elucidated the connection between the near-horizon CQM modes and the conventional Boulware modes for both the generalized Schwarzschild and Kerr geometries. These generalizations cover all black solutions in 4D and a variety of solutions in higher dimensionalities, thus
displaying the robustness of the framework. Possible future extensions of this work include understanding the connection of the HBAR entropy with the BH entropy and exploring the group-theoretical aspects of this problem.

\acknowledgments{}
M.O.S.\ and A.A.\ acknowledge support by the
National Science Foundation (Grant No.\ PHY-2013771), the Air Force Office of Scientific Research (Award No.\ FA9550-20-1-0366 DEF), the Office of Naval Research (Award No.\ N00014-20-1-2184), the Robert A. Welch Foundation (Grant No.\ A-1261) and the King Abdulaziz City for Science and Technology (KACST).
H.E.C. acknowledges support
by the University of San Francisco Faculty Development Fund. 
This material is based upon work supported by the Air Force Office of Scientific Research under award FA9550-21-1-0017 (C.R.O. and A.C.).  

\begin{appendix}
\section{Kerr geodesics}
\label{app:Kerr-geodesics}

 The geodesic equations can be reduced to their first-order form, written in terms of the conserved quantities:
 the energy $E= -p_{t} = -   \boldsymbol\xi_{(t)} \cdot {\bf p}$,
 the angular momentum along $z$-direction $L_z= p_{\phi}  = \boldsymbol\xi_{(\phi)} \cdot {\bf p}$
(as follows from the 4-momentum $ {\bf p}$ and the Killing vectors), the invariant mass $\mu$, and the Carter constant $\mathcal{Q}$. The geodesic equations are then given by
\begin{align}
\displaybreak[0]
\rho^2 \frac{dr}{d\tau} &= - \sqrt{\mathcal{R}(r)} \label{eq:geodesic1}\\
\displaybreak[0]
\rho^2 \frac{d\T}{d\tau} &= \pm\sqrt{\varTheta(\T)}\label{eq:geodesic2}\\
\displaybreak[0]
\rho^2 {\frac {d\phi }{d\tau }}
&=-\left(ae-{\frac {\ell}{\sin ^{2}\theta }}\right)+{\frac {a}{\Delta }}\mathcal{P}(r)\label{eq:geodesic3}\\ 
\displaybreak[0]
\rho^2 \frac{dt}{d\tau} &= -a(ae \s^2\T - \ell) + \frac{r^2+a^2}{\D} \mathcal{P}(r) \label{eq:geodesic4}
\; ,
\end{align}
with the auxiliary functions $\mathcal{P}(r),\,\mathcal{R}(r)$, and $\varTheta(\theta)$ defined by
\begin{align}
\displaybreak[0]
\mathcal{P}(r) &= e(r^2+a^2)-a\ell \label{eq:P} \\
\displaybreak[0]
\mathcal{R}(r) &= \left[ \mathcal{P}(r) \right]^2 - \D \left[r^2+(\ell-ae)^2+\mathfrak{q} \right] \label{eq:R} \\
\displaybreak[0]
\varTheta(\T) &= \mathfrak{q}-\C^2\T \left[a^2(1-e^2) + \frac{\ell^2}{\s^2\T}\right]    \label{eq:Theta}  \; .
\end{align}
Here, we have used specific (per unit mass) conserved quantities 
$e=E/\mu$, $\ell = L_z/\mu$, and $\mathfrak{q} = \mathcal{Q}/\mu = (p_{\theta }/\mu)^{2}+\cos ^{2}\theta \left[a^{2}\left( 1 - e^2 \right)+\left( {\ell}/{\sin \theta } \right)^{2}\right]$.  

In our approach, we only need the near-horizon behavior of the trajectories, which we extract with the hierarchical near-horizon expansion.
In terms of the near-horizon variable $x = r-r_+$. Using Taylor expansion around the horizon in terms of the variable $x$, the radial geodesic equation (\ref{eq:geodesic1}) becomes
 \begin{equation}
 \rho_+^2\frac{dx}{d\tau} 
 \stackrel{(\mathcal H)}{\sim} -\sqrt{c_0^2-c_1x + \mo( x^2) }\, ,
\label{eq:geodesic1-nh} 
\end{equation}
where $\rho_+^2 \equiv \rho_{+}^2 (\theta) = r_+^2 + a^2 \C^2\T$. 
The constants in Eq.~(\ref{eq:geodesic1-nh}) are functions 
of the conserved quantities of the motion and black hole parameters, and are given by 
\begin{align}
c_0 &=   \mathcal{P}(r_{+}) = \left( r_{+}^2 + a^2 \right) \left( e - \Omega_{H} \ell \right) 
=  \left( r_{+}^2 + a^2 \right)\,  \tilde{e}
 \label{eq:c_0} \\
c_1 &= -4e r_{+} c_{0} + \D'_{+}  \left[ r_+^2+(\ell - a e )^2 +\mathfrak{q} \right]
\label{eq:c_1} 
\; ,
\end{align}
where
\begin{equation}
\tilde{e} = -   \boldsymbol\xi_{(\tilde{t})} \cdot {\bf p}
= e - \Omega_{H} \ell > 0
\; 
\end{equation}
is the energy in a
frame rotating with the black hole's angular velocity $\Omega_{H}$.

However, we cannot yet integrate Eq.~(\ref{eq:geodesic1-nh}) to get the proper time in terms of $x$ due to its $\theta$ dependence through $\rho_+^2$, where $\theta$ evolves with the proper time following Eq.~(\ref{eq:geodesic2}). This issue can be resolved if there is a functional relationship between $\theta$ and $x$, independently of the other variables; this is indeed the case, as combining Eqs.~(\ref{eq:geodesic1}) and (\ref{eq:geodesic2}) yields the separable differential equation
\begin{equation}
    \frac{dr}{\sqrt{\mathcal{R}(r)}} = \mp  \frac{d \T }{\sqrt{\mathcal{\varTheta (\T)}}} 
    \; .
\end{equation}
Moreover, the near-horizon expansion and a change of variables $y\equiv\cos \theta$ reduce this equation 
to the form
\begin{equation}
    \frac{dx}{\sqrt{c_0^2-c_1x}} \stackrel{(\mathcal H)}{\sim}  \mp  \frac{d y}{\sqrt{\varTheta (y) \, (1-y^2)}}
    \label{eq:separable_theta-x}
    \; 
\end{equation}
[where, by abuse of notation, we have used $\varTheta (\theta)  \equiv \varTheta (y) $].
Equation~(\ref{eq:separable_theta-x}) provides a solution
$y( x)$ that can be written using elliptic integrals of the first kind; more general results can be found in Ref.~\cite{Chandrasekhar}. However, we can avoid using the full-fledged solution by extracting the leading-order near-horizon term, which gives
\begin{equation}
    y \stackrel{(\mathcal H)}{\sim} \text{const} + \mathcal{O}(x)\,.
\end{equation}
Reverting to the $\theta$ coordinate, we get $\cos \theta \stackrel{(\mathcal H)}{\sim} \cos\theta_+ + \mathcal{O}(x)$. Here $\theta_+$ is a constant value of the polar coordinate, which can be interpreted as the value of $\theta$ at which the atom approaches the event horizon. This subsequently leads to the replacement of $\rho_+^2 \rightarrow  \hat{\rho}_+^2 \equiv r_+^2 + a^2\cos^2\theta_+$. 
Any $\mathcal{O}(x)$ terms have been ignored because, after integration, they will yield additional contributions of order  $\mathcal{O}(x^2)$. Therefore, after making the replacement $\rho_+^2 \rightarrow  \hat{\rho}_+^2$, Eq.~(\ref{eq:geodesic1-nh}) can be integrated to give the proper-time functional relationship $\tau = \tau (x)$ to leading order,
\begin{equation}
\tau \stackrel{(\mathcal H)}{\sim} -k  x+ \mo(x^2) + \text{const} \; ,
\label{eq:tau-vs-x_app}
\end{equation}
in which $k=\hat{\rho}_+^2/c_0$. 
Furthermore, the factors $\rho^2$  (which are $\theta$-dependent) in
 Eqs.~(\ref{eq:geodesic3}) and (\ref{eq:geodesic4}) can be removed through division by 
Eq.~(\ref{eq:geodesic1}); this gives the functional relationships $t = t(x)$ and $\phi= \phi(x)$
by integration in the near-horizon limit,
\begin{align}
    t  &\stackrel{(\mathcal H)}{\sim} - \frac{1}{2 \kappa} \ln x - C x +\mo(x^2) \; , \label{eq:t-vs-x_app} \\
    \tilde{\phi} &\stackrel{(\mathcal H)}{\sim} \alpha x +\mo(x^2)\; .  \label{eq:phi-vs-x_app}
\end{align}
Instead of finding $\phi$, we have calculated $\tilde{\phi} = \phi-\Omega_H t$ (which can be obtained by combining solutions for $\phi$ and $t$), because the CQM modes in Eq.~(\ref{eq:CQM_modes}) explicitly depend on this corotating azimuthal variable. Most importantly, even though both $\phi$ and $t$ have logarithmic terms proportional to $\ln x$, these cancel out when combined into the locally well-defined coordinate $\tilde{\phi}$.
Finally, the constants $C$ and $\alpha$ can be computed by collecting all the $\mathcal{O}(x)$ terms arising from the functions on the right-hand side of Eqs.~(\ref{eq:geodesic1})--(\ref{eq:geodesic4}); a straightforward  calculation gives
\begin{equation}
C = \frac{1}{2 \kappa} 
\left[ \frac{1}{2} \frac{c_{1}}{c_{0}^{2}} + 
\frac{2 r_{+}}{R_{+}^2} \frac{\left( \tilde{e} + \Omega_H \ell \right) }{\tilde{e}}
- \frac{1}{2 \kappa R_{+}^{2}}
- \frac{\Omega_{H}}{\tilde{e}} \left( a e s_{+}^2 - \ell \right)
\right]
\label{eq:constant-C_Kerr}
\end{equation}
and
\begin{equation}
\alpha
=
\Omega_{H} \frac{r_{+}}{\kappa R_{+}^2}
- \frac{ \left( a e s_{+}^2 - \ell \right) \left( \Omega_{H} a - 1/ s_{+}^2 \right) }{R_{+}^2 \tilde{e}}
\; ,
\label{eq:constant-alpha}
\end{equation}
where $R_{+}^2 = r_{+}^2 + a^2$ and $s_{+} = \sin \theta_{+}$.
The constants $C$, $\alpha$, and $k$, as shown in Sec.~\ref{sec:conformal_steady_state}, do not play a direct role in the radiation formulas.

\section{Boulware vacuum and field modes for the Kerr black hole}
\label{app:vacuum_modes}
In this appendix, we address the issue of the nonuniqueness of a Boulware vacuum for the Kerr metric.
Specifically, for this metric, when the asymptotic-infinity regions are included in the spacetime manifold,
the presence of superradiant modes precludes the existence of a unique Boulware vacuum that is empty at both past ($\mathscr{I}^-$) and future ($\mathscr{I}^+$) null infinity---see the details that follow in this appendix. 
 It is noteworthy that, even though this is a technical issue that needs to be addressed, as shown below and in Sec.~\ref{sec:conformal_steady_state}, the use of a generic Boulware-like vacuum in our setup is sufficient---the final results for the acceleration radiation and the HBAR entropy are insensitive to whatever choice is made within this apparent ambiguity. 

We begin by discussing the conventional past ($B^{-}$) and a future ($B^{+}$) Boulware vacuum states
defined on the Cauchy surfaces $\mathscr{I}^-\cup\mathcal{H}^-$ and  $\mathscr{I}^+\cup\mathcal{H}^+$ respectively~\cite{frolov,ottewill00,menezes17}.
For comparison purposes with the existing literature, instead of using the near-horizon variables and CQM modes of our paper, one can describe the relevant physics with the equivalent tortoise coordinate $r_*$, defined by
 \begin{equation}
\frac{dr_*}{dr} = 
\frac{1}{f(r)}
\; \; , \; \; \; \text{where} \; \; \;
 f\equiv \frac{{\D}}{({r^2+a^2)}}
\label{eq:tortoisedef}
\; ,
\end{equation}
which implies that
$
    r_* = \int dr \,{r^2+a^2}/{\D} $.
This coordinate choice is made so that the radial-time sector of the metric appears as 
near-horizon conformally flat and pushes the horizon radially to minus infinity. Notice that the scale factor $f(r)$ plays the same role as the homologous factor in generalized Schwarzschild coordinates.
In the corotating coordinates~(\ref{eq:rotating-coords_separation_2}), the radial function $R(r)$ satisfies the wave equation
\begin{equation}
\left[\frac{d^2}{dr_*^2}+\tilde{\omega}^2\right] R(r)=0
\label{eq:Kerr_conformal-tortoise}
\, .
\end{equation}
Most importantly,
Eq.~(\ref{eq:Kerr_conformal-tortoise}) with the tortoise coordinate is equivalent to its counterpart
with the regular Boyer-Lindquist radial variable~(\ref{eq:Kerr_Klein_Gordon_conformal}).
The ingoing and outgoing waves 
$\left\{ e^{-i\tilde{\omega}(\tilde{t}+r_*)},e^{-i\tilde{\omega}(\tilde{t}-r_*)}\right\}$ in terms of $r_*$ 
correspond to the conformal ingoing/outgoing modes $x^{\mp i \Theta}$ of CQM.
More generally (for all regions of spacertime), the coordinate transformation of Eq.~(\ref{eq:tortoisedef}) converts the radial equation into the equivalent form
\begin{equation}
    \left[\frac{d^2}{dr_*^2} - V(r_*)\right] R(r) = 0
    \; ,
\end{equation}
with $r_* \in (-\infty,\infty)$, the event horizon being located at $r_* =-\infty$, and the scattering problem
taking a conventional form with 
an asymptotic potential $V(r_*) \equiv V_{\omega l m}(r_*)$ given by
\begin{equation}
    V(r_*) \sim \begin{cases}
    -\tilde{\omega}^2 \hskip 3em r_*\rightarrow -\infty\;\; (r\rightarrow r_+)\\
    -\omega^2 \hskip 3em r_*\rightarrow \infty\;\; (r\rightarrow \infty)
    \end{cases}\,. \label{eq:tortoise_pot_asymptotic}
\end{equation}
This leads to the following sets of asymptotic solutions
\begin{equation}
    R^-_{\omega l m} \sim \begin{cases}
    e^{i\tilde{\omega}r_*} + A^-_{\omega l m} e^{-i\tilde{\omega}r_*} \hskip 3em r_*\rightarrow -\infty\\
    B^-_{\omega l m} e^{i\omega r_*} \hskip 6.5em r_*\rightarrow \infty 
    \end{cases}\;, \label{eq:Rminus_modes}
\end{equation}
\begin{equation}
    R^+_{\omega l m} \sim \begin{cases}
    B^+_{\omega l m} e^{-i\tilde{\omega}r_*} \hskip 6.5em r_*\rightarrow -\infty\\
    e^{-i\omega r_*} + A^+_{\omega l m} e^{i\omega r_*} \hskip 3em r_*\rightarrow \infty
    \end{cases}\;. \label{eq:Rplus_modes}
\end{equation}
For the definition of properly normalized modes, we are adopting a convention that is standard in the literature, where, in the separation Eqs.~(\ref{eq:KG_separation}) and (\ref{eq:rotating-coords_separation_2}), and in the radial equation~(\ref{eq:Kerr_Klein_Gordon_radial}), we make the replacement
$R \longrightarrow (r^2+a^2)^{-1/2}R$; the corresponding normalized expressions are shown below.
The labeling of the modes follows the notation ${\boldsymbol{s}} = (\omega,l,m)$, of Sec.~\ref{sec:setup}.
It should be noted that the modes near the horizon $r_*\rightarrow -\infty$ are characterized in terms of $\tilde{\omega}$, since the positive frequency is defined with respect to the Killing vector $\xi_{(\tilde{t})}$ near the horizon (i.e., of the form  $\xi_{(\tilde{t})} \phi_{\boldsymbol{s}}  = - i \tilde{\omega} \phi_{\boldsymbol{s}}$);
whereas at asymptotic infinity the relevant Killing vector is $\partial_t$, which defines the positive frequency modes in terms of $\omega$. The $A$ and $B$ coefficients can be thought of reflection and transmission amplitudes from the potential barrier $V(r_*)$. One can show that for $\omega>0,\, \tilde{\omega}<0$, both $|A^-|^2$ and $|A^+|^2$ are greater than one, which is the phenomenon of superradiance. This peculiar behavior only arises for corotating waves ($m>0$), in the frequency range $-m\Omega_H<\tilde{\omega}<0$.

Using these modes, we define the past Boulware basis, with Cauchy data on the past surface $\mathscr{I}^-\cup\mathcal{H}^-$. The {\it in} modes are defined as waves coming from $\mathscr{I}^-$, transmitted to $\mathcal{H}^+$, and reflected to $\mathscr{I}^+$, with zero flux coming from $\mathcal{H}^-$. The {\it up} modes are defined as waves coming from $\mathcal{H}^-$, transmitted to $\mathscr{I}^+$, and reflected to $\mathcal{H}^+$, with zero flux coming from $\mathscr{I}^-$.
In other words, these are complementary modes that have unit flux and zero flux in the corresponding portions of the past surface 
 $\mathscr{I}^-\cup\mathcal{H}^-$. Specifically,
\begin{align}
    \phi^{\rm in}_{\omega l m} &= \frac{1}{\sqrt{8\pi^2\omega(r^2+a^2)}} e^{-i\omega t} e^{im\phi} S_{\omega l m}(\theta) R^+_{\omega l m}(r) \hskip 6em \omega>0 \;, \label{eq:in_modes}\\
    \phi^{\rm up}_{\omega l m} &= \frac{1}{\sqrt{8\pi^2\tilde{\omega}(r^2+a^2)}} e^{-i\tilde{\omega} t} e^{im\tilde{\phi}} S_{\omega l m}(\theta) R^-_{\omega l m}(r) \hskip 6em \tilde{\omega}>0 \;, \label{eq:up_modes}\\
    \phi^{\rm up}_{-\omega l -m} &= \frac{1}{\sqrt{8\pi^2(-\tilde{\omega})(r^2+a^2)}} e^{i\tilde{\omega} t} e^{-im\tilde{\phi}} S_{\omega l m}(\theta) R^-_{-\omega l -m}(r) \hskip 3em -m\Omega_H<\tilde{\omega}<0 \;. \label{eq:up_sup_modes}
\end{align}
Accordingly, the field can be quantized using these orthonormal-basis modes,
\begin{equation}
\Phi = \sum_{l,m} \int_0^\infty  d\omega \left(a^{\rm in}_{\omega l m} \phi^{\rm in}_{\omega l m} + {\rm h.c.} \right)  + \int_0^\infty  d\tilde{\omega} \left(a^{\rm up}_{\omega l m} \phi^{\rm up}_{\omega l m} + {\rm h.c.} \right) \;. \label{eq:past_quantization}
\end{equation}
The corresponding past Boulware vacuum is then defined by $a^\Lambda_{\boldsymbol{s}} \ket{B^-} = 0$, where $\Lambda \in \{\rm in,\,up\}$ and ${\boldsymbol{s}} = \{\omega, l, m\}$ or $\{-\omega, l, -m\}$. The past Boulware vacuum implies that there are no particles in the past.

Similarly, one can define future Boulware modes by using Cauchy data in the future surface $\mathscr{I}^+\cup\mathcal{H}^+$, with {\it out} and {\it down} modes. The {\it out} modes are defined as waves that reach only $\mathscr{I}^+$ from the past, with zero flux going into $\mathcal{H}^+$; and the {\it down} modes
are defined as waves that reach only $\mathcal{H}^+$ from the past, with zero flux going into $\mathscr{I}^+$.  In other words, these are complementary modes that have unit flux and zero flux in the corresponding portions of the future surface $\mathscr{I}^+\cup\mathcal{H}^+$. Thus, the out-down modes are the time-reversed versions of the in-up modes, and can be obtained via the complex conjugate of the radial wave functions; specifically,
\begin{align}
    \phi^{\rm out}_{\omega l m} &= \frac{1}{\sqrt{8\pi^2\omega(r^2+a^2)}} e^{-i\omega t} e^{im\phi} S_{\omega l m}(\theta) R^{+\;*}_{\omega l m}(r) \hskip 6em \omega>0\;, \label{eq:out_modes}\\
    \phi^{\rm down}_{\omega l m} &= \frac{1}{\sqrt{8\pi^2\tilde{\omega}(r^2+a^2)}} e^{-i\tilde{\omega} t} e^{im\tilde{\phi}} S_{\omega l m}(\theta) R^{-\;*}_{\omega l m}(r) \hskip 6em \tilde{\omega}>0 \;, \label{eq:down_modes}\\
    \phi^{\rm down}_{-\omega l -m} &= \frac{1}{\sqrt{8\pi^2(-\tilde{\omega})(r^2+a^2)}} e^{i\tilde{\omega} t} e^{-im\tilde{\phi}} S_{\omega l m}(\theta) R^{-\;*}_{-\omega l -m}(r) \hskip 3em -m\Omega_H<\tilde{\omega}<0 \;. \label{eq:dwon_sup_modes}
\end{align}
Similarly to Eq.~(\ref{eq:past_quantization}), these future orthonormal-basis modes can be used to quantize the scalar field as
\begin{equation}
    \Phi = \sum_{l,m} \int_0^\infty  d\omega \left(a^{\rm out}_{\omega l m} \phi^{\rm out}_{\omega l m} + {\rm h.c.} \right)  + \int_0^\infty  d\tilde{\omega} \left(a^{\rm down}_{\omega l m} u^{\rm down}_{\omega l m} + {\rm h.c.} \right)\;. \label{eq:future_quantization}
\end{equation}
We can define the corresponding future Boulware vacuum as $a^\Lambda_{\boldsymbol{s}} \ket{B^+} = 0$, where $\Lambda \in \{\rm out,\,down\}$ and ${\boldsymbol{s}} = \{\omega, l, m\}$ or $\{-\omega, l, -m\}$, which implies that there are no particles in the future. The main lesson of this construction is that the past and future vacuum states are not the same due to the existence of nontrivial Bogoliubov coefficients for the superradiant modes.

Having discussed the conventional vacuum modes for the Kerr geometry, we come back to the near-horizon ingoing and outgoing CQM modes given in Eq.~(\ref{eq:CQM_modes}). One can use the near-horizon expansion of the tortoise coordinate of Eq.~(\ref{eq:tortoisedef}) to show that the near-horizon CQM modes [Eq.~(\ref{eq:CQM_modes})] have the following correspondence with the in and out modes
\begin{equation}
\phi^{\mathrm{out}}_{\omega l m}
\stackrel{(\mathcal H)}{\propto} 
\Phi^{ + {\rm \scriptscriptstyle (CQM)} }_{\boldsymbol{s}} 
\;  , \; \; \;  \; \; \; 
\phi^{\mathrm{in}}_{\omega l m}
\stackrel{(\mathcal H)}{\propto} 
\Phi^{ - {\rm \scriptscriptstyle (CQM)} }_{\boldsymbol{s}} 
\; ,
\label{eq:CQM-vs-global-modes}
\end{equation}
whereas the up and down modes include contributions from both $\Phi_{\boldsymbol{s}}^{\pm{\rm \scriptscriptstyle (CQM)}}$ modes. 
The correspondence~(\ref{eq:CQM-vs-global-modes}) validates the use of a generic Boulware-like vacuum. For any such a vacuum, the ingoing/outgoing CQM waves directly yield a Planck distribution for the non-superradiant modes, as shown in Sec.~\ref{sec:conformal_steady_state}.
Choosing the future Boulware vacuum state has an apparent technical advantage in that the Planck function already includes the superradiant modes, but the past Boulware vacuum could be used just as well. 

Alternatively, it is also possible to define a Boulware vacuum by isolating the near-horizon behavior
responsible for the dominant conformal physics . This can be achieved by 
redefining the vacuum in the presence of an outer mirror or boundary condition located within the speed-of-light surface~\cite{rotating-Qvacuum}, as discussed in Sec.~\ref{sec:conformal_steady_state}.

\end{appendix}


\end{document}